\documentclass[11pt,letterpaper,nofootinbib]{revtex4-2}
\usepackage[top=100pt,bottom=0pt,letterpaper]{geometry}
\usepackage{adjustbox}
\usepackage[T1]{fontenc}
\usepackage{lmodern}
\usepackage[utf8]{inputenc}
\usepackage{microtype}
\usepackage{setspace}
\usepackage{float}
\restylefloat{table}

\usepackage{dcolumn}
\usepackage{lipsum}
\usepackage{graphicx,epsfig}
\usepackage{psfrag}
\usepackage{bm}
\usepackage{epstopdf}
\usepackage{latexsym}
\usepackage{multirow}
\usepackage{amsmath,amssymb}
\usepackage{enumerate}
\usepackage{hyperref}
\usepackage[FIGTOPCAP]{subfigure}
\usepackage{empheq,mathtools}
\usepackage{xcolor}
\usepackage{booktabs}
\usepackage{tabularx}
\usepackage{longtable}
\def\be {\begin{equation}}
\def\ee {\end{equation}}
\def\ba {\begin{eqnarray}}
\def\ea {\end{eqnarray}}
\def\nn {\nonumber}
\def\bc {\begin{center}}
\def\ec {\end{center}}
\newcommand{\bdm}{\begin{displaymath}}
\newcommand{\edm}{\end{displaymath}}

\def\l  {\lambda}

\def\nn {\nonumber}
\def\ra {\rightarrow}

\def\la {\label}
\def\le {\left}
\def\ri {\right}

\def\f {\frac}

\def\bi {\begin{itemize}}
\def\ei {\end{itemize}}

\def\> {\rangle}
\def\< {\langle}

\def\bc {\begin{center}}
\def\ec {\end{center}}

\usepackage{amsfonts}

\begin{document}
\title{Geodesics in Generalised Ellis-Bronnikov Spacetime Embedded in Warped 5D Background}

\author{Vivek Sharma} \email[email: ]{svivek829@gmail.com}
\affiliation{Department of Physics, Indira Gandhi National Tribal University, Amarkantak, M.P-484887, India}
\author{Suman Ghosh} \email[email: ]{suman.ghosh@bitmesra.ac.in}
\affiliation{Department of Physics, Birla Institute of Technology, Mesra, Ranchi-835215, India}

\begin{abstract}

We study the particle trajectories in the recently proposed five dimensional warped (generalized) Ellis-Bronnikov spacetime (5D-WGEB) (which does not require exotic matter) as well as it's four dimensional counterpart (4D-GEB) in detail and conduct a comparative study. Analytical approach provides conditions which determines three types of trajectories (trapped, returning and crossing) for both 4D and 5D spacetimes. Notably in 5D geometry existence of trapped trajectories become highly sensitive to the initial conditions.
We have solved the timelike geodesic equations numerically and presented the trajectories graphically along with corresponding geodesic potentials. We thus distinguished the roles of the so-called wormhole parameter and the warping factor regarding their effects on the trajectories and embedding as such. Interestingly, the 5D-WGEB model shows the localization of massive particles around the location of the brane for growing warp factor and runaway trajectories for decaying warp factor.

\end{abstract}


\maketitle
\section{introduction}

Wormholes are solutions of Einstein's field equations that connect two separate points within our world (intra-Universe wormholes) or two distinct points of parallel Universes (inter-Universe wormholes) \citep{Visser:1995cc, Lobo-book}. In 1935, Einstein and Rosen introduced the Einstein-Rosen Bridge, a regular solution that connects two asymptotically flat regions of spacetime and is a special case of the Schwarzschild black hole \cite{Einstein:1935tc}. Wheeler referred to these solutions as ``wormholes'' \citep{Wheeler:1957mu}. Further, it was shown that these wormholes are {\em non-traversable} \citep{Kruskal:1959vx, Fuller:1962zza, Eardley:1974zz, Wald:1980nm}. In general, {\em traversable} wormholes require violation of the so-called energy conditions to prevent the collapse of it's throat and this can be met by threading ``exotic matter'' (matter with negative energy density) at least at the throat \cite{Morris:1988cz}. Exotic forms of matter are popular in cosmology because of their benefits \cite{Lobo:2005us}, such as explaining the universe's accelerated expansion. However, it has also been pointed out that the methods involving quantum aspects of standard model matter is not adequate  to create  macroscopic wormholes \cite{Witten:2019qhl}.

Despite the facts mentioned previously, there are some classical methods developed in order to avoid the need for matter with negative energy density that violates energy conditions \cite{Hochberg:1990is, Bhawal:1992sz, Agnese:1995kd, Samanta:2018hbw, Fukutaka:1989zb, Ghoroku:1992tz, Furey:2004rq, Bronnikov:2009az}. There are alternative theories of gravity or modified gravity theories that provide new techniques to circumvent the violation of energy conditions. A significant number of non-exotic matter models under modified gravity can be found in the literature \cite{Lobo:2008zu, Kanti:2011jz, Kanti:2011yv, Zubair:2017oir, Shaikh:2016dpl, Ovgun:2018xys, Canate:2019spb,Myrzakulov:2015kda}, albeit the convergence condition of null geodesics is violated in some circumstances. The dynamical wormhole models also provide ways to keep wormhole throat open in the presence of a viable matter source \citep{Hochberg:1998ii, Roman:1992xj, Kar:1994tz, Kar:1995ss, Visser:2003yf}. The so-called $f (R)$, $f(R, T)$, $f(Q)$ and higher order gravity theories are other popular class modified gravity theories that involve comprehensive examination of wormhole geometries with feasible matter sources \citep{Lobo:2009ip, Garcia:2010xb, MontelongoGarcia:2010xd, Sajadi:2011oei, Moraes:2017dbs, Sahoo:2017ual, Moraes:2019pao, Sahoo:2020sva, Hassan:2021egb, Mustafa:2021ykn}.

Although wormholes are still considered hypothetical, recent advances in precision measurements related to black holes have increased the significance for testing viable wormhole models (as black hole mimicker) as well.
Studies on various phenomena, such as wormhole merger  \citep{Krishnendu:2017shb, Cardoso:2016oxy} or their quasinormal modes \citep{Aneesh:2018hlp, DuttaRoy:2019hij}, can be beneficial for capturing wormhole signatures in the cosmos. In principle, one can also detect them through their lensing effects, shadows, einstein ring etc. \citep{Abe:2010ap, Toki:2011zu, Takahashi:2013jqa, Cramer:1994qj, Perlick:2003vg, Tsukamoto:2012xs, Bambi:2013nla, Nedkova:2013msa, Zhou:2016koy, Dzhunushaliev:2011xx,  Dzhunushaliev:2012ke, Dzhunushaliev:2013lna, Dzhunushaliev:2014mza, Aringazin:2014rva, Dzhunushaliev:2016ylj,Vagnozzi:2022moj}. Interestingly, signatures like these may also favour the case for modified gravity theories over general relativity.

One of the most studied traversable wormhole geometry is the four dimensional Ellis-Bronnikov spacetime (4D-EB) \citep{Ellis:1973yv, Bronnikov:1973fh},  which is sustained by a phantom scalar field (a field with a negative kinetic term). Various aspects for this spacetime has been studied in the context of general relativity (GR) and also modified gravity theory such as- geometry of spinning 4D-EB spacetime \cite{Chew:2016epf}, generalized spining of 4D-EB wormhole in scalar-tensor theory \cite{Chew:2018vjp}, hairy Ellis wormholes solutions \cite{Chew:2020svi}, Ellis Wormholes in Anti-De Sitter Space \cite{Blazquez-Salcedo:2020nsa}, stability analysis of 4D-EB solution in higher dimensional spacetime \cite{Torii:2013xba} etc. Kar et al. 
To avoid violation of energy conditions \cite{Kar:1995jz} constructed a generalized version of the 4D-EB wormhole (4D-GEB)  by introducing a new wormhole parameter $m \ge 2$ ($m=2$ case represents the original 4D-EB model). Recently, they also looked at the quasi-normal modes, echoes, and other aspects of this spacetime \cite{DuttaRoy:2019hij}. Motivated by this model we recently proposed a model where the 4D-GEB geometry is embedded in five dimensional warped spacetime discussed below. 

The theories of extra dimensions appear in fundamental physics quite naturally. In fact, it dates back a century, when Kaluza (1921) and Klein (1926) attempted to merge electromagnetism and gravity in a 5D gravity model \cite{Kaluza:1921tu, Klein:1926tv}. 
Extra spatial dimensions were reinvented through unification models (such as superstrings) \cite{Green:1987sp}. A warped extra dimension can solve the age-old Hiararchy problem \cite{Rubakov:1983bb, Gogberashvili:1998vx}. Extra dimensions are also an essential ingredient in the octonionic thoeries of the standard model particle physics (and what may lay beyond) \cite{Furey:2015yxg, Baez:2001dm, Baez:2010ye, Furey:2018yyy, Furey:2018drh, Gillard:2019ygk}. 
The so-called `warped braneworld' models \citep{Gogberashvili:1998iu, Randall:1999ee, Randall:1999vf} are probably the most well-known of these higher-dimensional models. This model posits a non-factorizable geometry – a curved five-dimensional spacetime in which the 4D-metric depends the extra dimension via a warping factor (a feature unique to this class of models). Though there are some investigations reported recently on wormholes embedded in higher-dimensional spacetime 
 \citep{Lobo:2007qi, deLeon:2009pu, Wong:2011pt, Kar:2015lma, Banerjee:2019ssy, Wang:2017cnd}, warped braneworld models have not been considered as such.
  
In \cite{Sharma:2021kqb}, we demonstrated that a warped GEB (5D-WGEB) model, where a generalized version of Ellis-Bronnikov spacetime is embedded in a 5D warped background, satisfies the energy conditions, even for $m=2$. Therefore, in this work also we shall focus mostly on the original E-B geometry embedded in 5D warped background (5D-WEB).
Note that, the 5D line element we used is the well-known thick braneworld model \citep{Dzhunushaliev:2009va, Ghosh:2008vc}, in which the warp factor 
is a smooth function of extra spatial dimension (unlike the Randall-Sundrum model). Hence derivative jumps and delta functions do not occur in the curvature and connections (which is one of the reasons to work with this model). 
For a decaying warp factor, the matter source satisfies the weak energy condition and violates the strong energy condition. Opposite feature is observed in presence of a growing warp factor. Thus one can say that the 5D wormhole model can be supported by bulk normal matter field for decaying warp factor and exotic matter field for growing warp factor.
In this article, we investigate geodesics in detail for both the 4D-GEB and 5D-WGEB model and compare them in order to distinguish the role of the wormhole parameter and warped extra dimension. 

This paper is organized as follows. The wormhole spacetimes corresponding to our 5D model is introduced in Section \ref{sec:wg}. It contains a brief description of the Wormhole characteristics and the warping factor and embedding diagrams for both wormhole models. 
In Section \ref{sec:geo}, through analytic approach (wherever possible), dynamical systems analysis and embedding diagrams we built intuitions about the trajectories and corresponding effective potentials. 
In Section \ref{sec:num}, we solve the geodesic equations for various initial conditions numerically and presented the geodetic potentials and particle trajectories for both 4D-GEB and 5D-WGEB geometries.  Finally, we compare the 4D and 5D models based on our key findings and provide conclusion of this study with a summary of the results in Section \ref{sec:dis}.


\section{Wormhole-Geometry} \label{sec:wg}
The general metric for warped 5D spacetime can be written as,
\begin{equation}
ds^{2} = e^{2f(y)} g_{\mu \nu}~dx^{\mu}~dx^{\nu} + g_{44}~dy^{2} \label{eq:5d-general-line-element}
\end{equation}

where, $g_{\mu \nu}$ is any 4D metric and $g_{44}$ can be a function of 3-Space, time and extra spatial dimension $y$ ($ - \infty \leq y \leq \infty$), not necessarily separable. We choose a specific metric for  our 5D-WGEB spacetime as follows,
\begin{equation}
ds^{2} =  e^{2f(y)} \Big[ - dt^{2} +  dl^{2} + r^{2}(l)~\big(  d\theta^{2} + \sin^{2}(\theta)~d\phi^{2} \big) \Big] + dy^{2} \label{eq:5d-line-element}
\end{equation}
where, $f(y)$ is the warp factor (we choose $f(y) = \pm \log[\cosh(y/y_{0})]$ that represents a thick domain wall in the 5D bulk peaked at $y=0$) and the term in the square bracket is the 4D spherically symmetric, ultra-static wormhole model, called the Generalised Ellis-Bronnikov space-time and is given by,
\begin{equation}
ds^{2}_{4D} = - dt^{2} +  dl^{2} + r^{2}(l)~\big(  d\theta^{2} + \sin^{2}(\theta)~d\phi^{2} \big) \label{eq:generalised-EB-l}
\end{equation}
\begin{equation}
\mbox{with}~~~~~ r(l) = (b_{0}^{m} + l^{m})^{1/m}. \label{eq:r(l)}
\end{equation}
Here $l$ is the `proper radial distance' or `tortoise coordinate'. $b_{0}$ is the so-called `throat radius' of the wormhole and $m$ is the wormhole parameter that can take only even values ($m \geq 2$) (to ensure the smooth behaviour of $r(l)$). Note that, metric (\ref{eq:generalised-EB-l}) can also be written in usual radial coordinate $r$ as 
\begin{equation}
ds^{2} = - dt^{2} + \frac{dr^{2}}{\Big( 1 - \frac{b(r)}{r} \Big)} + r^{2} \big( d\theta^{2} + \sin^{2}\theta d\phi^{2} \big), \label{eq:generalised-EB-r}
\end{equation}
where $r$ and $l$ are related through the {\em shape function} $b(r)$ as,
\begin{equation}
dl^{2} = \frac{dr^{2}}{ \Big( 1 - \frac{b(r)}{r} \Big)}~~~~~\implies ~~~~~\quad b(r) = r - r^{(3-2m)} (r^{m} - b_{0}^{m})^{ \Big( 2 - \frac{2}{m} \Big)}. \label{eq:r-l-relation}
\end{equation} 
One gets back the Ellis-Bronniokv geometry, for $m = 2$, which is a static, spherically symmetric, geodesically complete, horizonless space-time (constructed using phantom scalar field) represented by the 4D metric as given below, 
\begin{equation}
ds^{2} = - dt^{2} + \frac{dr^{2}}{ \Big( 1 - \frac{b_{0}}{r} \Big) } + r^{2} \big( d\theta^{2} + \sin^{2}\theta d\phi^{2} \big). \label{eq:EB-metric}
\end{equation}
The Ricchi Scalar ($R_{5D}$) and Kretschmann scalar ($K_{5D}$) for the metric \ref{eq:5d-line-element} is given by
\be
R_{5D} = 2e^{-2f}\le(\f{(2m-3)l^{2m-2}}{(b_0^m + l^m)^2} - \f{2(m-1)l^{m-2}}{b_0^m + l^m} + \f{1}{(b_0^m + l^m)^{2/m}}\ri) -4 (5f'^2 + 2 f'') ,\la{eq:Ricci-5D}
\ee
\begin{eqnarray}
& &  K_{5D} = 4 \left[ 3 f'(y)^{4} +  \frac{2 e^{-4f(y)} \left( b_{0}^{m} l^{m} (-1+m) + e^{2f(y)} l^{2} \left( b_{0}^{m} + l^{m} \right)^{2} f'(y)^{2} \right)^{2}}{l^{4} \left( b_{0}^{m} + l^{m}  \right)^{4}} +  \right. \nn\\
& &  \left.   \frac{e^{-4f(y)}} {( b_{0}^{m} + l^{m} )^{4/m}} \left( -1 + \f{l^{-2+2m}} {b_{0}^{m} + l^{m})^{2 - \frac{2}{m}}} + e^{2f(y)} \left( b_{0}^{m} + l^{m} \right)^{2/m} f'(y)^{2}  \right)^{2} + 4 \left( f'(y)^{2} + f''(y)  \right)^{2} \right] ~~~~
\label{eq:GEB-sigsq}
\end{eqnarray}
Thus the curvature invariants of our 5D model are essentially singularity free (i.e. they do not show any divergence at any finite values of the coordinates or at the `throat' as such) unlike some models of black holes in higher dimensions.
Asymptotically  ($y \rightarrow \pm \infty$), the Ricci scalar gives negative constant value ($-20$) for growing warp factor and large positive value for decaying warp factor which says that the asymptotic regions (along $y$) are not flat. However the four dimensional wormhole passage is asymptotically (at $l \rightarrow \pm \infty$) flat.
Before addressing the geodesic equations, let us discuss the isometric embedding of the wormhole which provides a useful perspective on the geometry of these exotic objects.

\subsection{Isometric-Embedding}

\begin{figure}[H]
\centering
\includegraphics[scale=.78]{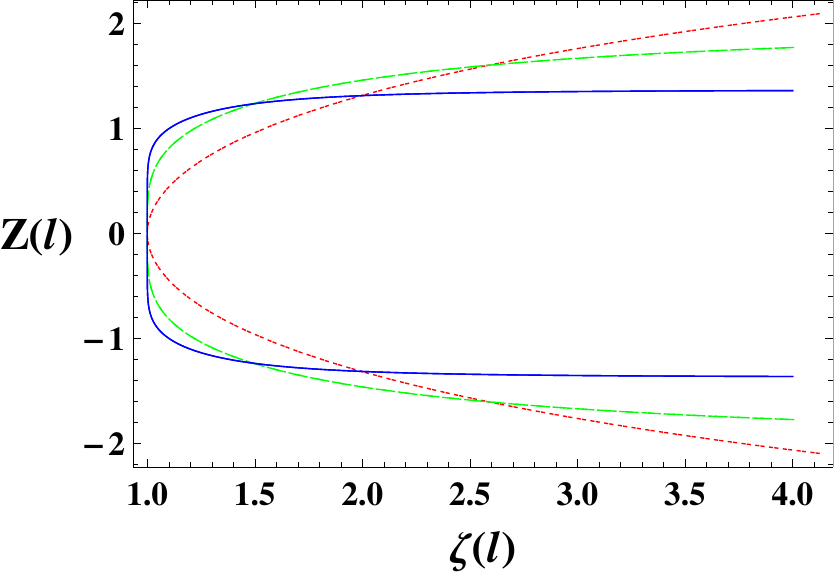}
\caption{Embedding of 4D-GEB wormhole geometry for different choice of ``$m$'' with $b_{0} = 1$, where cases; $m = 2, 4, 8$ are represented by dotted-red, dashed-green and solid-blue curves, respectively.}
\label{fig:EB-embedding}
\end{figure}

\begin{figure}[H]
\centering
\includegraphics[scale=.5]{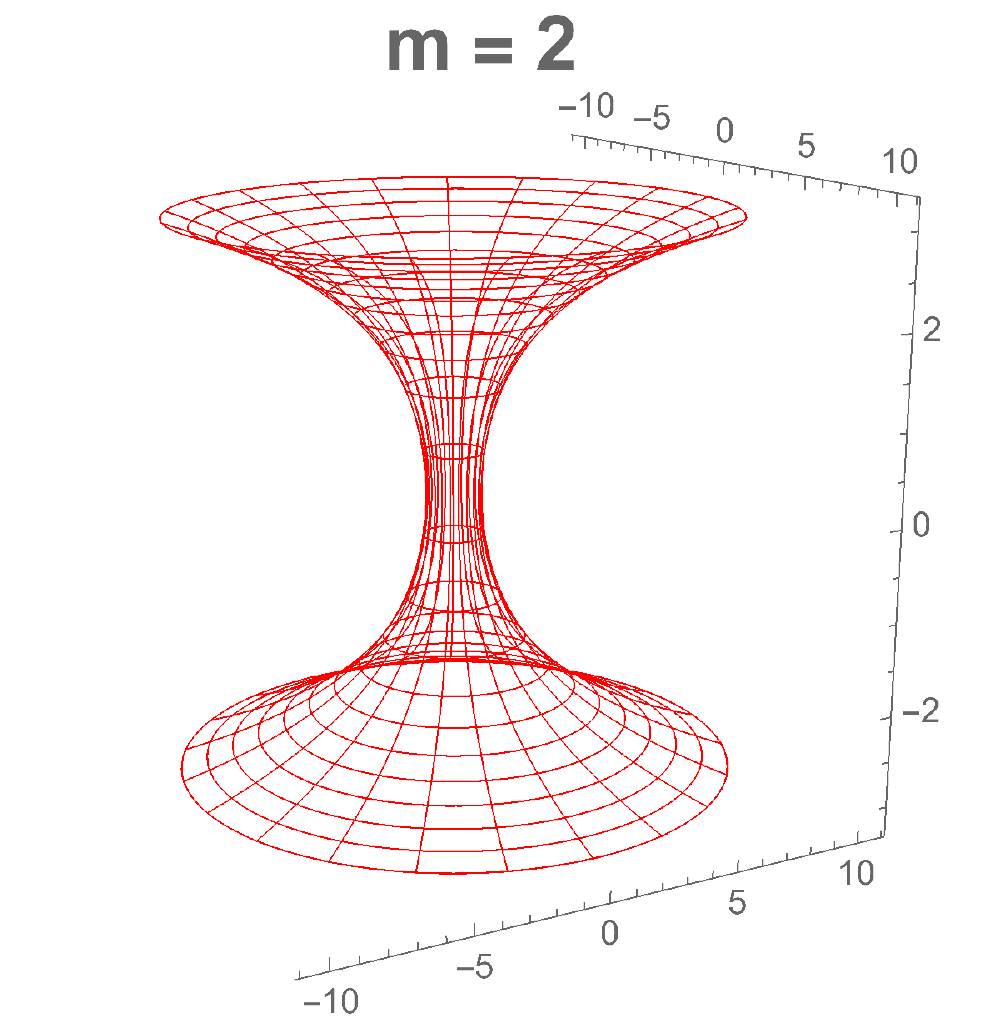}
\hspace{1cm}
\includegraphics[scale=.48]{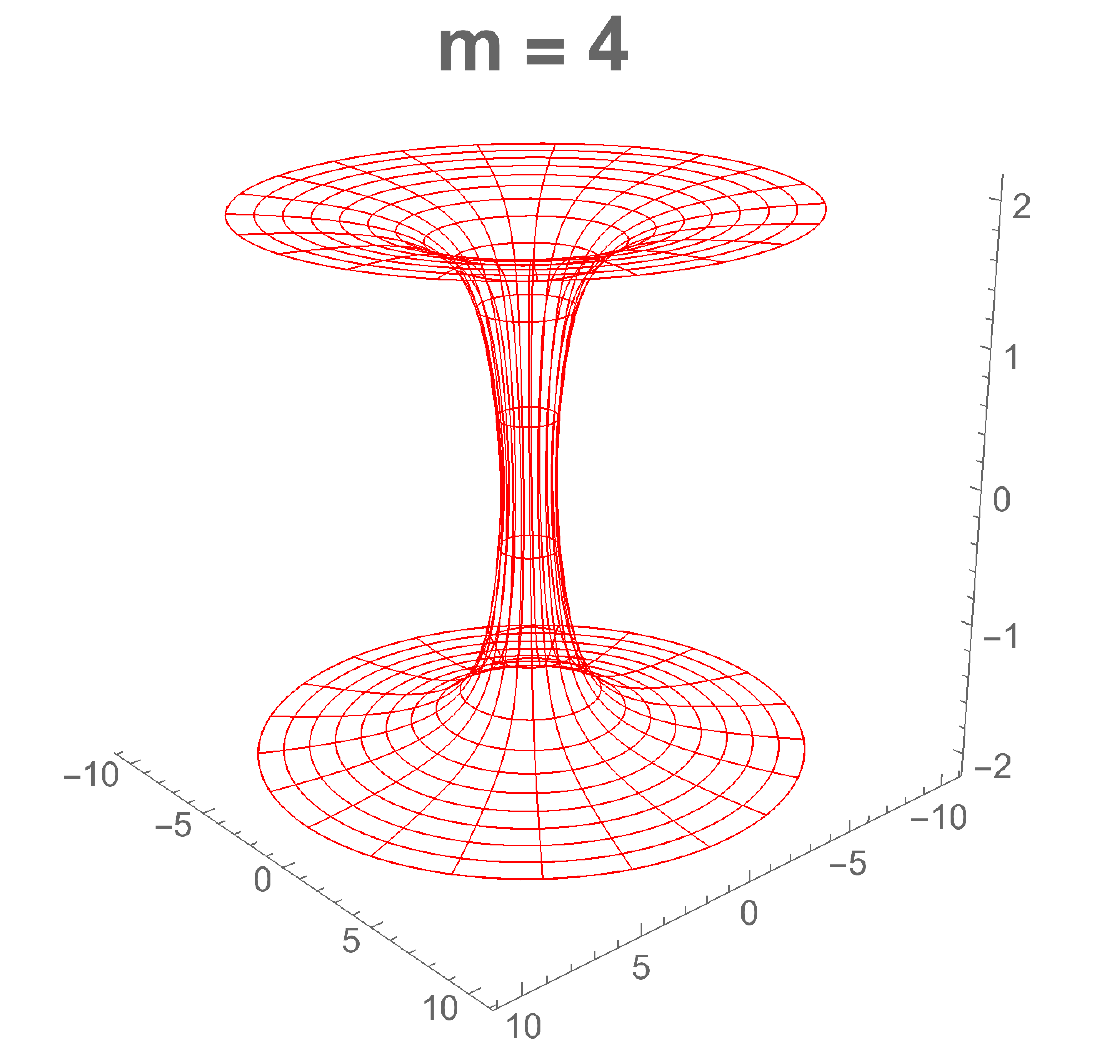}
\caption{Surface of revolution of embedding for 4D-GEB model.}
\label{fig:EB-sor}
\end{figure}

Embedding of lower dimensional space-time in a flat space of higher dimension can be significant for various reasons, e.g., to visualise the `shape' of any general-space-time. In general, a $d$-dimensional Riemannian-space can be immersed in a flat space of $\frac{d(d+1)}{2}$ dimension. Such a $d$-dimensional Riemannian-metric is said to be of `embedding class $p$', if and only if, it can be be embedded in a flat space of (lowest possible) dimension $d+p$. it is well known that the general spherically symmetric space-time is of `class two' but, if any spherical symmetric space-time will satisfy the `Karmarkar condition' (KMC) then it is of class one \citep{Kuhfittig:2021sph, karmarkar-1948} as such. KMc for any general 4D spherically symmetric spacetime is given by,
\begin{equation}
R_{1414} = \frac{ R_{1212}R_{3434} + R_{1224} R_{1334} }{ R_{2323} } \label{eq:KMc}
\end{equation}
where, $ R_{2323} \neq 0 $. One can easily check that the condition mentioned above is satisfied for 4D-GEB wormhole geometry \ref{eq:generalised-EB-r}. 
The 2D spatial slice ($t$ = constant, $\theta = \frac{\pi}{2}$) for 4D-GEB spacetime is given by,
\begin{equation}
ds^{2} = dl^{2} + (b_{0}^{m} + l^{m})^{2/m} d\phi^{2}. \label{eq:2d-slice-4d} 
\end{equation}
One may immerse this 2D geometry in a 3D Euclidean space. Since Eq. (\ref{eq:2d-slice-4d}) possesses axial symmetry, we take a 3D line-element of flat space in cylindrical coordinate ($\zeta, \psi, z$) given by,
\begin{equation}
d\sigma^{2} = d\zeta^{2} + \zeta^{2} d\psi^{2} + dZ^{2}. \label{eq:3d-line-element}
\end{equation}
Then we identify, $\psi = \phi$, $\zeta = \zeta(l)$ and $Z = Z(l)$ by comparing Eq. (\ref{eq:2d-slice-4d}) and (\ref{eq:3d-line-element}) which implies,
\begin{equation}
d\sigma^{2} = \Big[ \Big( \frac{d\zeta}{dl} \Big)^{2} + \Big( \frac{dZ}{dl} \Big)^{2} \Big] dl^{2} + \zeta^{2} d\phi^{2} \label{eq:3d-line-element2}
\end{equation}
\begin{equation}
\mbox{where} ~~~~~~~~ \zeta(l) = (b_{0}^{m} + l^{m})^{1/m} \label{eq:zeta(l)-4d}
\end{equation}
\begin{equation}
\mbox{and} ~~~~~~~~~~ \frac{dZ}{dl}  = \Big( 1 - l^{-2+2m} \big( b_{0}^{m} + l^{m} \big)^{-2 + \frac{2}{m}} \Big)^{1/2}. \label{eq:dz/dl-4d}
\end{equation}
Note that the right hand side of the Eq. (\ref{eq:dz/dl-5d}) goes to zero as $l \rightarrow \pm \infty$.
We integrate Eq. (\ref{eq:dz/dl-4d}) numerically with $b_{0} = 1$ and for $m = 2, 4$ and $8$. Fig. (\ref{fig:EB-embedding}) is a parametric plot of $Z(l)$ vs $\zeta(l)$ showing isometric embedding diagrams for different choice of wormhole parameter $m$. Throat radius ($l = 0$), lie on surface $Z = 0$. The GEB wormhole geometry differs for different values of $m$. The `neck length' of wormhole increases (or it becomes steeper) with increasing $m$.  Rotating  these $Z(l)$ vs $\zeta(l)$ plots around the $Z$ axis, as illustrated in Fig. (\ref{fig:EB-sor}), we get a so-called embedded surface visualisation as such.
It is well known that shape of the geometry corresponding to $m = 2$ (EB) embedded in 3D Euclidean-space is that of a catenoid (minimal surface; the surface with zero mean curvature) formed by rotation of catenary (mathematically catenary is a graph of the hyperbolic cosine function) about $Z$ axis. Note that higher order diagrams are different compared to the $m=2$ case essentially because at the throat, $\f{d^nZ}{dl^n}$ vanishes for $n=0,2,...,m$.

\subsection{Embedding of 5D-WGEB Model}

\begin{figure}[H]
\centering
\includegraphics[scale=.6]{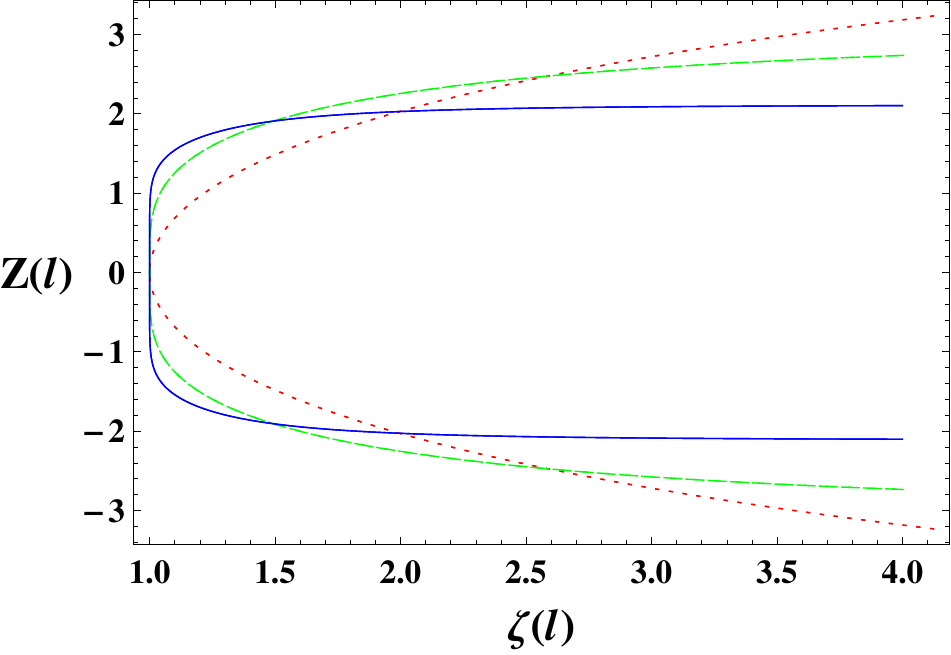}
\hspace{1cm}
\includegraphics[scale=.58]{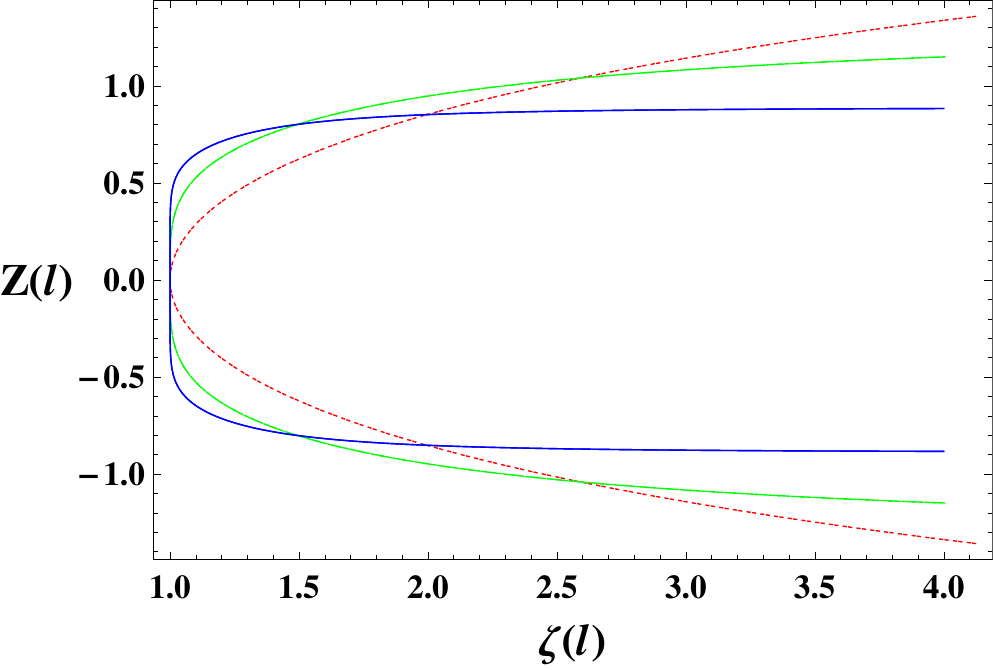}
\caption{Embedding diagrams of 5D-WGEB wormhole with growing (left) and decaying (right) warp-factor for  $m = 2, 4, 8$ represented by the dotted, dashed and continuous curves, respectively. }
\label{fig:5d-embedding-diagram}
\end{figure}

Similarly, we can construct an embedding for our 5D-WGEB model (using counterpart of Eq. (\ref{eq:dz/dl-4d})) as well. The 2D spatial slice (for $t = y = constant, \theta = \frac{\pi}{2}$) for 5D-WGEB spacetime is,  
\begin{equation}
ds^{2} = e^{2f(y)} \Big(dl^{2} + (b_{0}^{m} + l^{m})^{2/m} d\phi^{2} \Big).\label{eq:2d-slice-5d} 
\end{equation}
We write $\psi = \phi$, $\zeta = \zeta(l)$ and $Z = Z(l)$as before. 
Comparing the Eqs. (\ref{eq:3d-line-element2}) and (\ref{eq:2d-slice-5d}) we get,  
\begin{equation}
\zeta(l) = e^{f(y)} (b_{0}^{m} + l^{m})^{1/m} \label{eq:zeta(l)-5d}
\end{equation}
\begin{equation}
 \frac{dZ}{dl}  = e^{f(y)} \Big( 1 - l^{-2+2m} \big( b_{0}^{m} + l^{m} \big)^{-2 + \frac{2}{m}} \Big)^{1/2} \label{eq:dz/dl-5d}
\end{equation}
Note that, the rate of change of $Z(l)$ increases (decreases), for growing warp-factor (for decaying warp factor), with increasing $y$ (i.e. as one moves farther away into the extra dimension). As before, we integrate the Eq. (\ref{eq:dz/dl-5d}), numerically, for finite value of $y= y_0 = 1$ say (for both growing and decaying warp factors) and for $m=2,4,8$. The parametric plots of $Z(l)$ vs $\zeta(l)$ are presented in Fig. (\ref{fig:5d-embedding-diagram}). It is clear that the presence of warp factor affect the neck length of wormholes considerably.
Compared to the 4D case, the neck-length in general increases (decreases) for growing (decaying) warp factor.
Away from $y=0$, with increasing $y$, the neck-length is larger for growing warp-factor compared to the same in presence of the decaying warp-factor. 
Note that the extra dimension do effect the topology of the wormhole as the wormhole geometry is now five dimensional. In the 4D-EB or 4D-GEB scenario, the boundary or the asymptotic regions are topologically trivial i.e. geometrically flat whereas in our 5D-WEB model, due to the warping factor, flat boundary regions are absent as discussed above.
In the 4D scenarios, on a spatial slice, the radius of the throat is given by $r=b_0$ or $l=0$ which essentially is a 2-sphere $S^2$. In our 5D model, this 4D wormhole is embedded in a 5D bulk and the point $l=0$ represent a $S^2 \otimes R^1$ space. However, the particles that are confined near $y=0$, experiences the same throat radius. The volume of the `throat' of an effective 4D wormhole passage is scaled by $e^{2f(y)}$ as a result of warping.
Let us now go on to the detailed analysis of the timelike geodesics.


\section{Geodesics}  \label{sec:geo}

The free-falling test particles obey the geodesic equations given by,
\begin{equation}
\frac{d^{2}x^{\mu}}{d\lambda^{2}} + \Gamma^{\mu}_{\rho\sigma} \frac{dx^{\rho}}{d\lambda} \frac{dx^{\sigma}}{d\lambda} = 0 \label{eq:G-E}
\end{equation}
where, $x^{\mu}$ are coordinates, 
 $\lambda$ is the affine parameter and $ \Gamma^{\mu}_{\rho\sigma}$ are the affine connections.  
The geodesic equations for the 4D GEB spacetimes are given by,
\begin{equation}
\frac{d^{2}t}{d\lambda^{2}} = 0 \label{eq:geodesic-1}
\end{equation}
\begin{equation}
\frac{d^{2}l}{d\lambda^{2}} - l^{-1+m}  ~\big( b_{0}^{m} + l^{m} \big)^{-1 + \frac{2}{m}} ~\Big[ \Big( \frac{d\theta}{d\lambda} \Big)^{2} + \sin^{2}\theta  ~\Big( \frac{d\phi}{d\lambda} \Big)^{2} \Big] = 0 \label{eq:geodesic-2}
\end{equation}
\begin{equation}
\frac{d^{2}\theta}{d\lambda^{2}} + \frac{2l^{-1+m}}{\big( b_{0}^{m} + l^{m} \big)} \frac{dl}{d\lambda} \frac{d\theta}{d\lambda} - \sin\theta \cos\theta \Big( \frac{d\phi}{d\lambda} \Big)^{2} = 0 \label{eq:geodesic-3}
\end{equation}
\begin{equation}
\frac{d^{2}\phi}{d\lambda^{2}} + 2 \cot\theta \frac{d\theta}{d\lambda} \frac{d\phi}{d\lambda} + \frac{2l^{-1+m}}{ \big( b_{0}^{m} + l^{m} \big) } \frac{dl}{d\lambda} \frac{d\phi}{d\lambda} = 0 \label{eq:geodesic-4}
\end{equation}

The corresponding equaton for the 5D-WGEB model are as follows. 
\begin{equation}
\frac{d^{2}t}{d\lambda^{2}} + 2 ~f'(y) ~\frac{dt}{d\lambda} ~\frac{dy}{d\lambda} = 0 \label{eq:geodesic-11}
\end{equation}
\begin{equation}
\frac{d^{2}l}{d\lambda^{2}} + 2~f'(y)~\frac{dl}{d\lambda}~\frac{dy}{d\lambda}  - l^{-1+m}  ~\big( b_{0}^{m} + l^{m} \big)^{-1 + \frac{2}{m}} ~\Big[ \Big( \frac{d\theta}{d\lambda} \Big)^{2} + \sin^{2}\theta  ~\Big( \frac{d\phi}{d\lambda} \Big)^{2} \Big] = 0 \label{eq:geodesic-22}
\end{equation}
\begin{equation}
\frac{d^{2}\theta}{d\lambda^{2}} + 2 ~f'(y) ~\frac{d\theta}{d\lambda} ~\frac{dy}{d\lambda} + ~\frac{2l^{-1+m}}{(b_{0}^{m} + l^{m})} ~\frac{d\theta}{d\lambda} ~\frac{dl}{d\lambda} - \sin\theta ~\cos\theta ~\Big( \frac{d\phi}{d\lambda} \Big)^{2} = 0 \label{eq:geodesic-33} 
\end{equation}
\begin{equation}
\frac{d^{2}\phi}{d\lambda^{2}} + 2 ~f'(y) ~\frac{d\phi}{d\lambda} ~\frac{dy}{d\lambda} + 2 ~\cot\theta ~\frac{d\theta}{d\lambda} ~\frac{d\phi}{d\lambda} + \frac{2l^{-1+m}}{ \big( b_{0}^{m} + l^{m} \big) } ~\frac{dl}{d\lambda} ~\frac{d\phi}{d\lambda} = 0 \label{eq:geodesic-44}
\end{equation}
\begin{equation}
\frac{d^{2}y}{d\lambda^{2}} + f'(y)~e^{2f(y)}~\Big[ \Big( \frac{dt}{d\lambda} \Big)^{2} - \Big( \frac{dl}{d\lambda} \Big)^{2} - (b_{0}^{m} + l^{m})^{2/m} ~\Big[ \Big( \frac{d\theta}{d\lambda} \Big)^{2} + \sin^{2}\theta ~\Big( \frac{d\phi}{d\lambda} \Big)^{2} \Big] \Big] = 0 \label{eq:geodesic-55}
\end{equation}
At a glance, the difference between 4D and 5D geodesic equations is that there are extra terms involving $\dot{y}$ on the right hand side of Eqs. (\ref{eq:geodesic-11} - \ref{eq:geodesic-44}) and an extra geodesic equation for motion along the fifth dimension $y$. Given the difficulty in solving the full geodesic equations analytically, we shall use various analytic approximation as well as numerical approaches to understand the complete behaviour of the geodesics, which will be discussed in the subsequent sections.


\subsection{Analytical Approach}

Using the constants of the motion and the geodesic constraint one can reduce the highly coupled geodesic equations in simpler form in order to investigate key aspects of particle trajectories in the context of both the 4D GEB and 5D WGEB models. The geodesic constraint is essentially, 
\begin{equation}
g_{\mu \nu}u^{\mu}u^{\nu} = \epsilon, \label{eq:general-geodesic-constraint}
\end{equation}
where $u^{\mu}$ represents a four-velocity vector field and $\epsilon = -1, 0$ and $1$ for the timelike, lightlike and spacelike trajectories respectively. 

\subsubsection{4D-GEB model}
The geodesic constraint equation for the 4D GEB wormhole geometry leads to,
\begin{equation}
- \dot{t}^{2} + \dot{l}^{2} + (b_{0}^{m} + l^{m})^{2/m} \big( \dot{\theta}^{2} + \sin^{2}\theta ~\dot{\phi}^{2} \big) - \epsilon = 0, \label{eq:geodesic-constraint-4d}
\end{equation}
where an overdot denotes derivative with respect to the affine parameter $\lambda$. 
From the metric itself, one can derive the constants of the motion using the corresponding Euler-Lagrange's equations. The constants of the motion corresponding to the cyclic coordinates ($t$ and $\phi$) for the metric (\ref{eq:generalised-EB-l}) are, say,
\begin{equation}
\dot{t} = k  ~~~~~ \mbox{and} ~~~~~\sin^{2}\theta~(b_{0}^{m} + l^{m})^{2/m}\dot{\phi} =  h ,\label{eq:constants-4d}
\end{equation}
where $k$ and $h$  are integration constant. These constants can be thought of as conserved energy and angular momentum for lightlike particles, and conserved energy and angular momentum per unit mass of the particle for timelike particles. To simplify the geodesic equation we look at the equatorial plane where $\theta = \pi/2$. Using Eqs. (\ref{eq:geodesic-constraint-4d} - \ref{eq:constants-4d}) we get the following relation
\begin{equation}
\Big(\frac{dl}{d\phi} \Big)^{2} = \frac{(b_{0}^{m} + l^{m})^{2/m}  \Big[(b_{0}^{m} + l^{m})^{2/m}  (k^{2} + \epsilon) - h^{2} \Big]}{h^{2}}.
\end{equation} \label{eq:simplified-geodesic-4d}
Clearly there exist trajectories for which $\dot{l} = \frac{dl}{d\phi} = 0$ at certain value of $l = l_c$ (say), then this so-called point of return $l_c$ is given by,
\begin{equation}
l_{c}^{(4D)} = \Big[ \Big(\frac{h}{\sqrt{k^{2} + \epsilon}} \Big)^{m} - ~b_{0}^{m} \Big]^{1/m} \label{eq:critical-l-4d}
\end{equation}
In the 4D GEB wormhole, depending on whether $l_{c}^{(4D)} = 0,$ positive or imaginary, one can characterise three types of trajectories: trapped at the throat, returns before reaching the throat and ones that cross the throat and reaches the other side. These trajectories correspond to the following three conditions respectively,
\begin{equation}
\begin{split}
h^{m} = (b_{0} \sqrt{k^{2} + \epsilon})^{m}  \implies \text{Trapped trajectories}  \\
h^{m} > ( b_{0} \sqrt{k^{2} + \epsilon} )^{m} \implies \text{Returning trajectories} \\
h^{m} < ( b_{0} \sqrt{k^{2} + \epsilon} )^{m} \implies \text{Crossing trajectories} \\
\end{split}  
\label{eq:4d-trajectory-condition}
\end{equation}
Note that in certain class of models \cite{Willenborg:2018zsv} (where there is some asymmetry between the two Universes) another kind of trajectory is possible where a particle crosses to the other side but eventually returns back. However, for the class of models we are discussing here, no such behaviour is found (as further confirmed by the phase-space analysis discussed later).
\subsubsection{5D-WGEB model}

The metric constraint in 5D-WGEB spacetime is given by,
\begin{equation}
e^{2f(y)} \big[ - \dot{t}^{2} + \dot{l}^{2} + (b_{0}^{m} + l^{m})^{2/m}~\big( \dot{\theta}^{2} +  \sin^{2} \theta~ \dot{\phi}^{2} \big) \big] + \dot{y}^{2} - \epsilon = 0  ,\label{eq:geodesic-constraint-5d}
\end{equation}
The constants of motion are
\begin{equation}
e^{2f(y)}~\dot{t} = T ,\label{eq:constant1-5d}
\end{equation} 
\begin{equation}
e^{2f(y)}~\sin^{2} \theta~( b_{0}^{m} + l^{m} )^{2/m}~\dot{\phi} = H, \label{eq:constant2-5d}
\end{equation}
where $T$ ($T^{2}$ is the kinetic energy term) and $H$ (angular momentum per unit mass) are integration constants. At $\theta = \pi/2$, Eqs. (\ref{eq:geodesic-constraint-5d} - \ref{eq:constant2-5d}) leads to
\begin{equation}
\Big(  \frac{dl}{d\phi} \Big)^{2} = \frac{ (b_{0}^{m} + l^{m})^{2/m} \Big[ (b_{0}^{m} + l^{m})^{2/m} (T^{2} + \epsilon ~e^{2f(y)}) - H^{2}  \Big]  -  \frac{H^{2}}{e^{2f(y)}} \Big(  \frac{dy}{d\phi} \Big)^{2} }{H^{2}} .\label{eq:simplified-geodesic-5d}
\end{equation}
The point of return or the critical length $l_{c}^{(5D)}$ for the 5D-WGEB spacetime is then
\begin{equation}
l_{c}^{(5D)} = \left[\left( \frac{H}{ \sqrt{T^{2} - e^{2f(y)}~\big( \dot{y}^{2} - \epsilon \big)} } \right)^{m} - b_{0}^{m} \right]^{1/m}. \label{eq:critical-l-5d}
\end{equation}
Similar to the 4D-case, Eq. (\ref{eq:critical-l-5d}) also suggests the three types of trajectories: trapped, returning and crossing which correspond to the following three conditions respectively:
\begin{equation}
\begin{split}
H^{m} = (b_{0} \sqrt{T^{2} - e^{2f(y)}~(\dot{y}^{2} - \epsilon})^{m}  \implies \text{Trapped} \\
H^{m} > (b_{0} \sqrt{T^{2} - e^{2f(y)}~(\dot{y}^{2} - \epsilon})^{m} \implies \text{Returning} \\
H^{m} < (b_{0} \sqrt{T^{2} - e^{2f(y)}~(\dot{y}^{2} - \epsilon})^{m} \implies \text{Crossing} \\
\end{split}  
\label{eq:5d-trajectory-condition}
\end{equation}
Note the appearance of the $y$-dependent terms which considerably changes the critical length and the determining conditions for various types of geodesics compared to the 4D model. 
Note that the second term in the square root ($e^{2f(y)}~(\dot{y}^{2} - \epsilon)$) evolves with the affine parameter unlike the 4D conditions. 
Suppose a particle satisfies the `trapped' condition at a given time (or $\l$) and moving towards the throat. However, in case of decaying warp factor (as we will see in the next section), the right hand side will eventually become greater than the left hand side and we shall have a returning trajectory at the end.
Below we discuss the geodetic potentials to have another perspective on the geodesics. 

\subsection{Effective Potential}

The concept of the effective potential comes from the fact that typically potential of any field explicitly depends on position with respect to source, and so, if one is free to write the energy of any kind in terms of distance with respect to source then it can be treated as potential (called effective potential).
Using the general form of the constraint and constants of the motion, one may get the expression of effective geodesic potential for any space-time in general theory of relativity.
Thus for the case of 4D-GEB wormhole, the effective geodetic potential corresponding to motion along $l$ (at $\theta = \pi/2$) is given by
\begin{equation}
V_{l}^{(4)}(\lambda) = - \frac{1}{2} \dot{l}^{2} = - \frac{1}{2} \Big[ k^{2} - \frac{h^{2}}{(b_{0}^{n} + l^{n})^{2/n}} + \epsilon \Big] \label{eq:effective-potential-4d}
\end{equation} 
Similarly, the effective geodetic potentials for our 5D-WGEB wormhole corresponding to motion along $l$ and $y$, are given by
\begin{equation}
V_{l}^{(5)}(\lambda) = -\frac{1}{2e^{2f(y)}}~\Big[ \epsilon + \frac{1}{e^{2f(y)}} \big( T^{2} - \frac{H^{2}}{(b_{0}^{m} + l^{m})^{2/m}} \big) - \dot{y}^{2} \Big] ,\label{eq:effective-potential-5d1}
\end{equation}
\begin{equation}
V_{y}(\lambda) = - \frac{1}{2}~\Big[ \epsilon + \frac{1}{e^{2f(y)}} \big( T^{2} - \frac{H^{2}}{(b_{0}^{m} + l^{m})^{2/m}} \big) - e^{2f(y)}~\dot{l}^{2}  \Big] \label{eq:effective-potential-5d2}
\end{equation}
The distinguishing features of effective potentials for the 4D and 5D cases are essentially the $y$ and $\dot{y}$ dependent terms on the right hand side of the above equations. To investigate these potentials further, we make parametric plots for $V_{l}(\lambda)$ vs $l(\lambda)$ (for both the cases; 4D and 5D) and $V_{y}(\lambda)$ vs $y(\lambda)$ where  $l(\lambda)$ and $y(\lambda) $ can be found by numerically solving corresponding geodesic equations. 
Before doing so let us look at the following dynamical system analysis of the geodesic equations in 4D.


\subsection{Dynamical Systems Analysis}

Note that the geodesic equations (\ref{eq:geodesic-1}-\ref{eq:geodesic-4}) for the 4D-GEB model, on equatorial plane, can be written as a set of first coupled differential equations as given below, 
\begin{equation}
\dot{t} = k \label{eq:first-order-t}
\end{equation}
\begin{equation}
\dot{\phi} =  \frac{h}{(b_{0}^{m} + l^{m})^{2/m}} \label{eq:first-order-phi}
\end{equation}
\begin{equation}
\dot{l} = Q \quad \mbox{and} \quad
\dot{Q} = \frac{l^{-1+m}h^{2}}{(b_{0}^{m} + l^{m})^{1+\frac{2}{m}}} \label{eq:coupled-system1}
\end{equation}
Note that the equation for the tortoise coordinate decouples from the other equations and forms a set of two coupled first order differential equations as given by Eq. \ref{eq:coupled-system1}.
For $m \geq 2$, these equations in general, represent non-linear coupled systems. However, for $m=2$, these equations are linear in $l$ in the limit $l \ra 0$ and 	given by
\be
\dot{l} = Q , ~~~~~\dot{Q} = l ,~~~~ \mbox{where}~~~  h^{2} = 1 =  b_{0}^{2}, ~~m = 2
\ee 
The solutions for this system are the so-called `exponential-like solutions' given by 
\begin{empheq}[left=\empheqlbrace]{align}
    \begin{rcases}
    l = c_{1} e^{\lambda} + c_{2} e^{-\lambda}  \\
    Q = c_{1} e^{\lambda} - c_{2} e^{-\lambda}
  \end{rcases}. \label{eq.general-solution}
\end{empheq}
The linearised system $(\dot{l}, \dot{Q}) = (Q,l)$ can be treated as an representation of a vector field on $(l, Q)$ plane (the `phase plane' or `solution space'). The solution space then shows collection of solutions with all possible initial conditions \cite{Strogatz-1994, Ghosh:2009ig}. The phase space portraits for both $m = 2$ and $4$ are shown in Fig. (\ref{fig:phase-plot}). In general, we see that, for $m = 4$, test particles will spend more time as they get closer to the throat than the case $m = 2$. We also found that for large values of $l$, both cases show similar pattern. However near the throat $l=0$ trajectories are less similar. Below we discuss the implications of these phase plots in detail.
\begin{figure}[H]
\centering
\includegraphics[scale=.25]{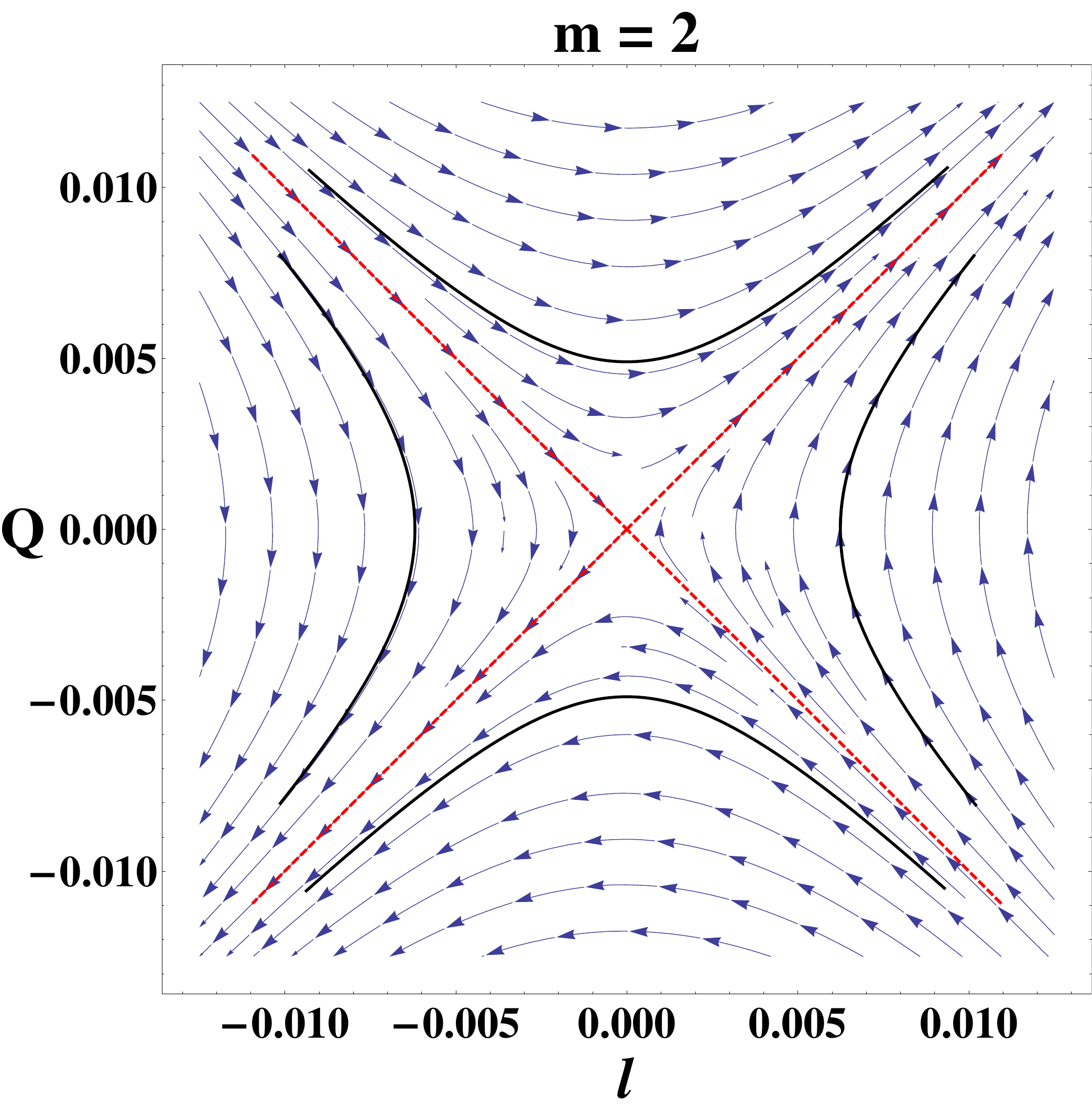}
\hspace{1cm}
\includegraphics[scale=.24]{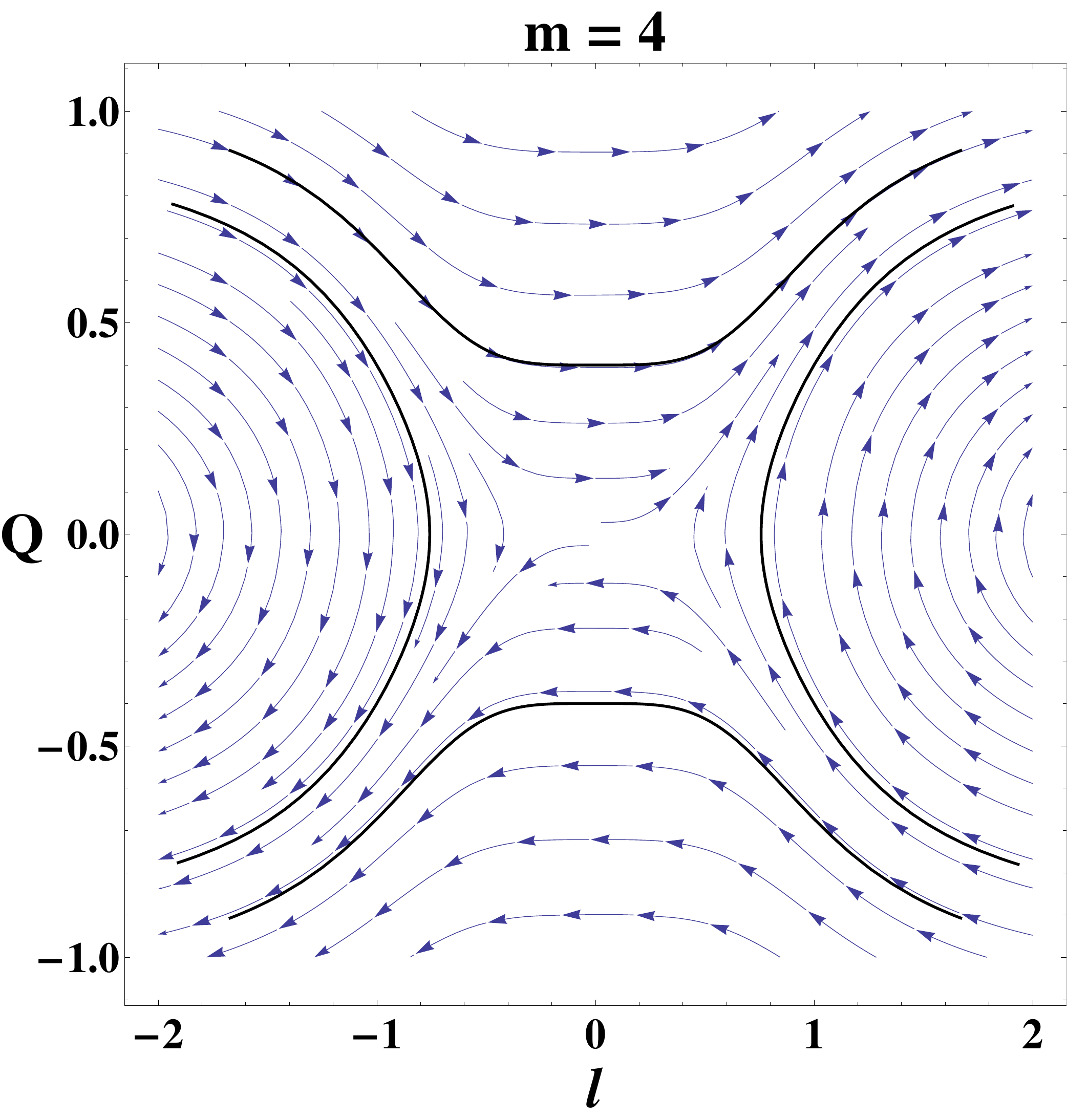}
\caption{Phase-space-plot for the coupled system corresponding to geodesic equation Eq.{eq:coupled-system1} for $m = 2, 4$). Here, we are considering $b_{0} = 1$ and $h = 1$.}
\la{fig:phase-plot}
\end{figure} 
Let us first look at the solution space of the linearised system for $m = 2$. Where, The point $(l, Q) = (0, 0)$ is a saddle point since the eigenvalues are real and product of the eigenvalues is negative. The trajectories on the straight line $l + Q = 0$ will end up at $(l, Q) = (0, 0)$ with an ever-decreasing velocity (trapped trajectories), but the trajectories on the straight line $l - Q = 0$ will fly out to infinity from the point $(l, Q) = (0, 0)$ with an ever-increasing velocity. This suggests instability under perturbation for the trapped trajectories at or near the throat. 
These two straight lines ($l + Q = 0$ and $l - Q = 0$) can be used to divide the phase plane into four quadrants; ($l + Q > 0$, $l - Q < 0 $: top quadrant), ($l + Q < 0$, $l - Q > 0$: bottom quadrant), ($l + Q > 0$, $l - Q > 0$: right quadrant) and ($ l + Q < 0 $, $ l - Q < 0$: left quadrant). 
The trajectories corresponding to the first two quadrants will cross the throat (in opposite direction) at some point of time depending on their initial conditions-- these are the crossing trajectories. The trajectories corresponding to the remaining quadrants will never cross the throat radius though they may come arbitrarily close depending on their initial conditions (returning trajectories). Thus, the dynamical systems analysis shows the presence of three types of trajectories (trapped, crossing and returning trajectories) for 4D-EB wormhole geometry, this analysis is difficult to perform for our 5D-WGEB model because the geodesic equations become highly coupled as such. So we solved the geodesic equations numerically for 5D-WGEB models and compare with that of 4D-GEB scenario in the next section. 

\section{Numerical Evaluations of the geodesic equations } \label{sec:num}

We numerically solved (using MATHEMATICA) the full geodesic equations (\ref{eq:geodesic-1}-\ref{eq:geodesic-55}) for both the 4D-GEB and 5D-WGEB (with growing and decaying warp-factor) wormhole geometries for the timelike trajectories ($\epsilon = -1 $). For 5D scenario, our focus will be on $m=2$ case (5D-WEB).
We have presented the trajectories and the corresponding effective geodetic potentials graphically, in the Figs (\ref{fig:4d-trajectories-m2}-\ref{fig:geodesics-effect-of-initial-y}). The boundary conditions, satisfying geodesic constraints and the conditions (\ref{eq:4d-trajectory-condition}) and (\ref{eq:5d-trajectory-condition}), used are listed in the Appendices. 

\subsection{4D-GEB model}

\begin{figure}[H]
\centering
\includegraphics[scale=.52]{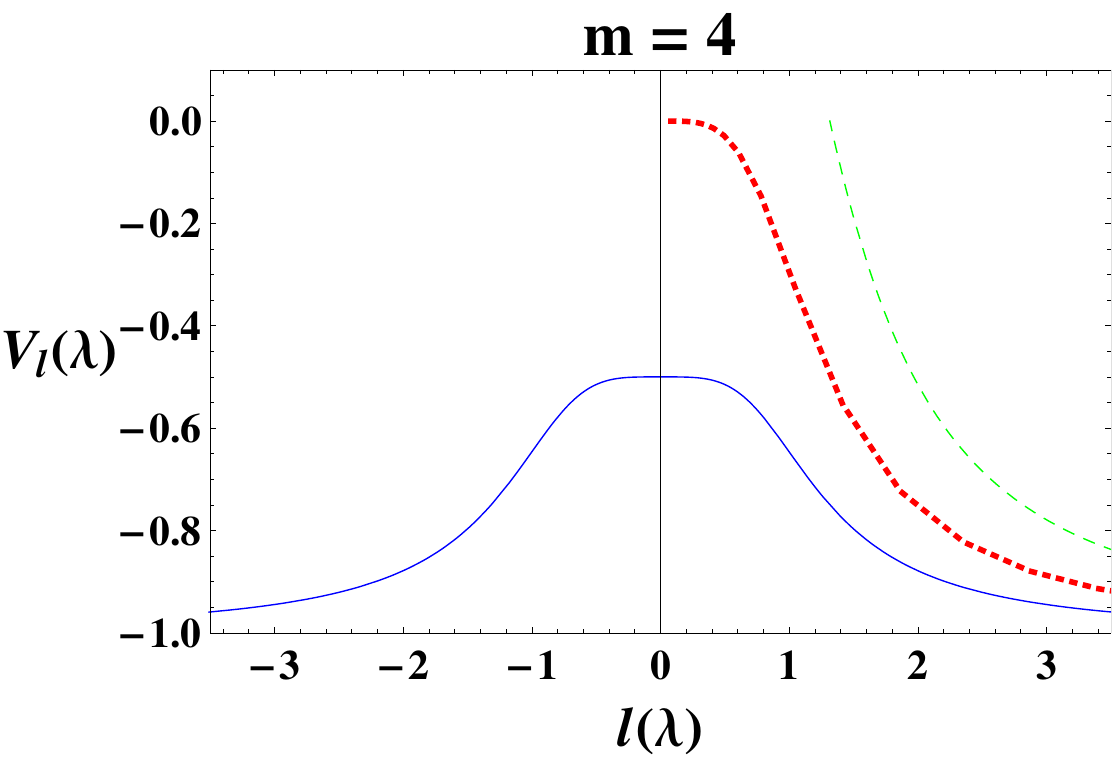}
\hspace{1 cm}
\includegraphics[scale=.5]{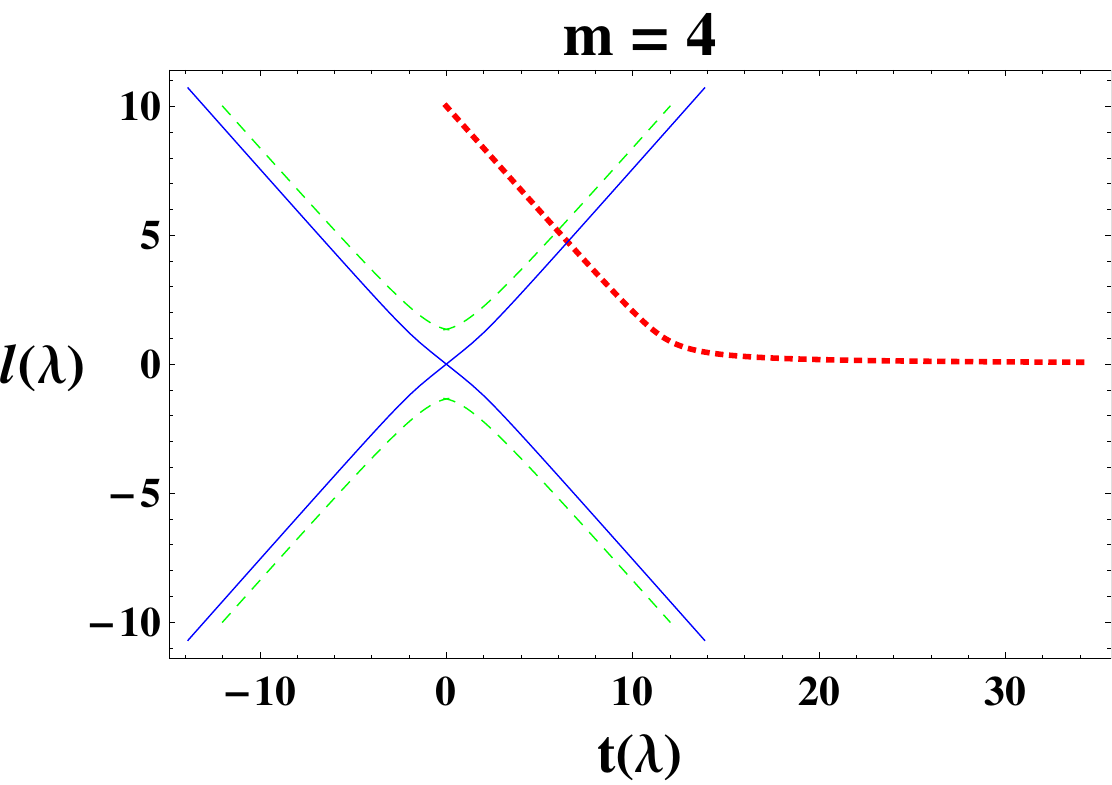}
\caption{Effective potential and geodesics for 4D-GEB wormhole ($m=4$): dotted-red, dashed-green, and continuous-blue curves representing -- Trapped, Returning and Crossing geodesics respectively.}
\label{fig:4d-trajectories-m4}
\end{figure}

\begin{figure}[H]
\centering
\includegraphics[scale=.52]{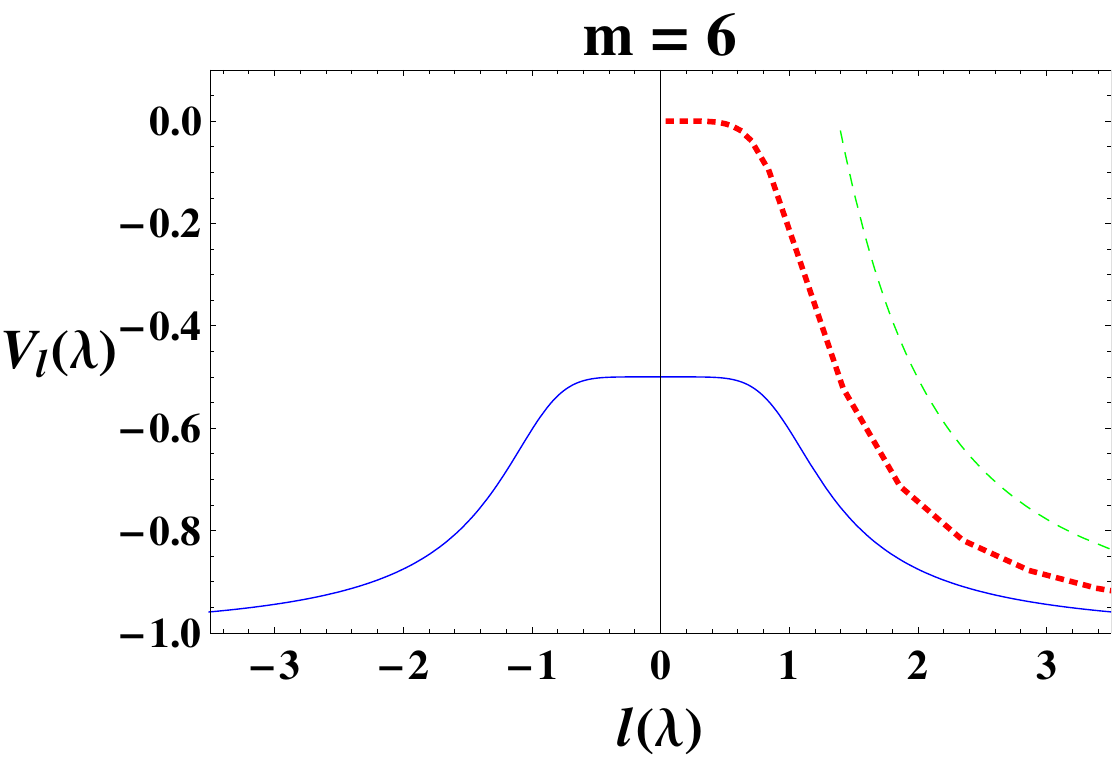}
\hspace{1cm}
\includegraphics[scale=.5]{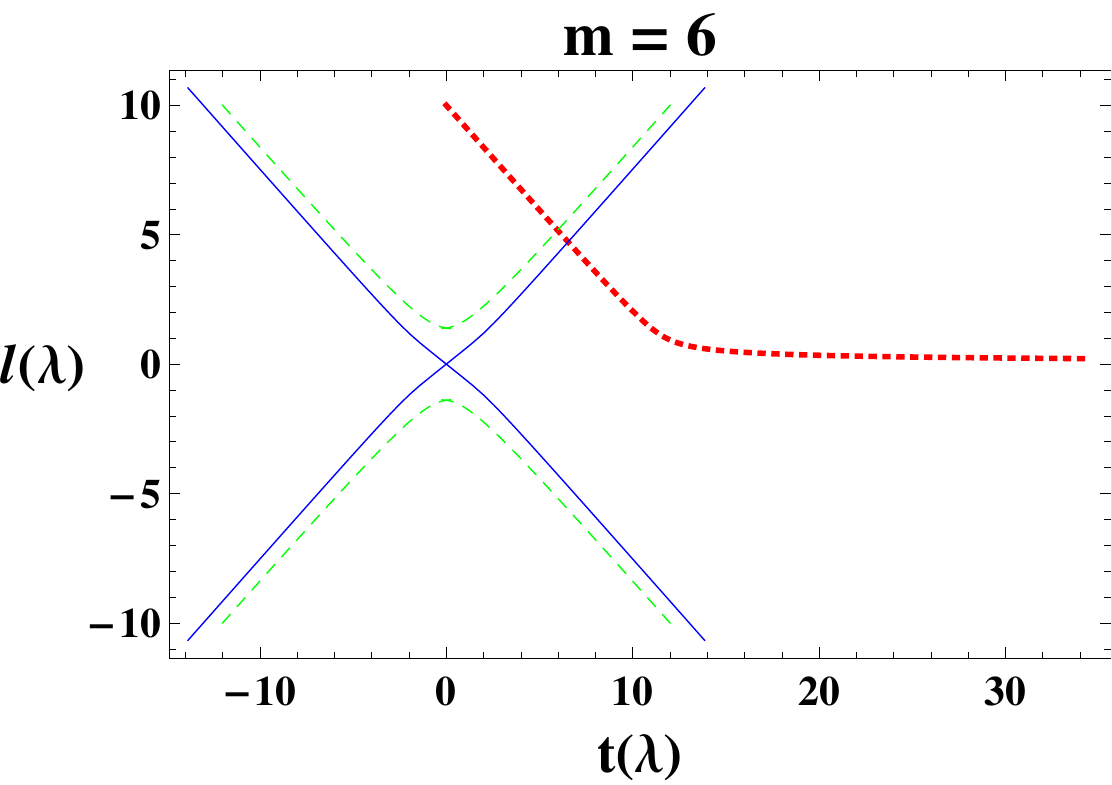}
\caption{Effective potential and geodesics for 4D-GEB wormhole ($m=6$): dotted-red, dashed-green, and continuous-blue curves representing -- Trapped, Returning and Crossing geodesics respectively.}
\label{fig:4d-trajectories-m6}
\end{figure}

Figures (\ref{fig:4d-trajectories-m2}-\ref{fig:4d-trajectories-m6}) show that for any value of the wormhole parameter $m$, there are three different class of geodesics that correspond to various boundary conditions (different energies and angular momentums as such). All the trajectories invariably slow down at the throat or near the point of return. The effective potentials, $V_{l}(\lambda)$, as $m$ is increased, flatten out at or around the throat ($l = 0$), implying that the tidal effect on the test particles will be reduced at or near the throat. This further suggests higher stability of particle trajectories for higher $m$. 
The consistency between the plots of the effective potential and the corresponding trajectory implies the accuracy of the numerical evaluation.
The so-called trapped trajectories are really trapped because at $l = 0$ both $\dot{l}$ and $\ddot{l}$ are zero. In other words, particles with angular momentum corresponding to trapped trajectories will take an indefinite time to spiral onto the throat. This can be seen be seen in the above figures (the red/dotted curves) as well. 
Thus full numerical analysis validates the three types of timelike trajectories: trapped, returning, and crossing trajectories for the 4D-GEB model as suggested by the analytic calculations in the previous section.
\begin{figure}[H]
\centering
\includegraphics[scale=.51]{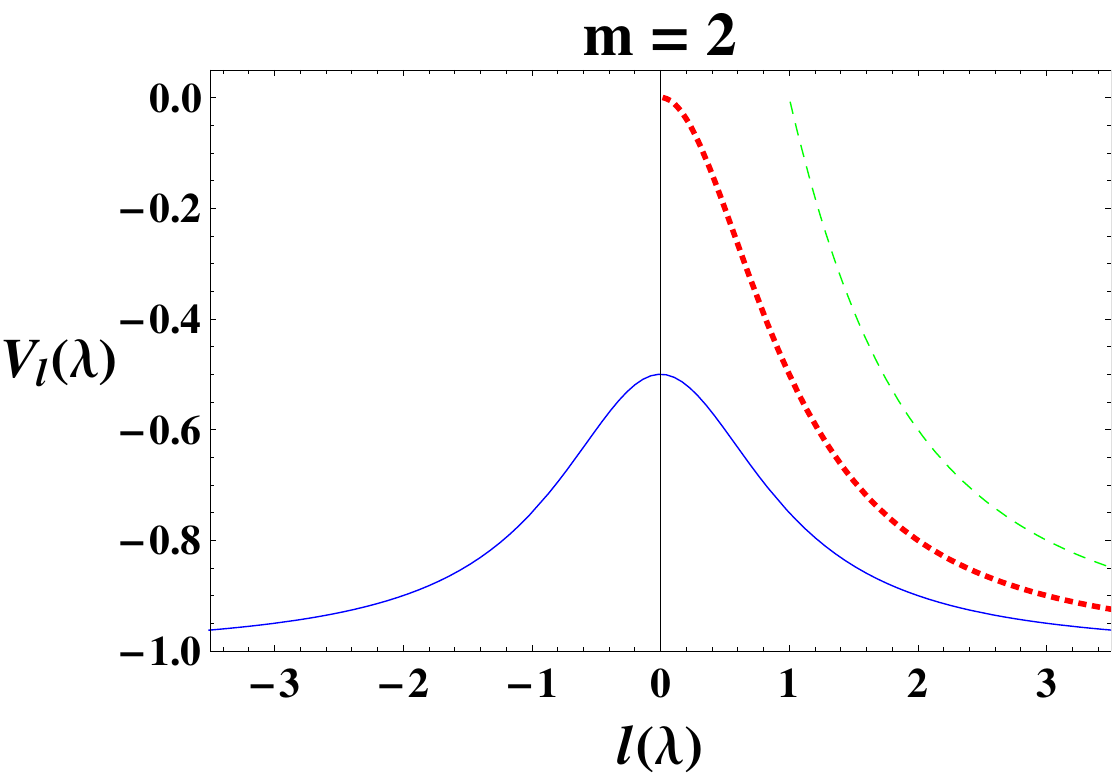}
\hspace{1cm}
\includegraphics[scale=.5]{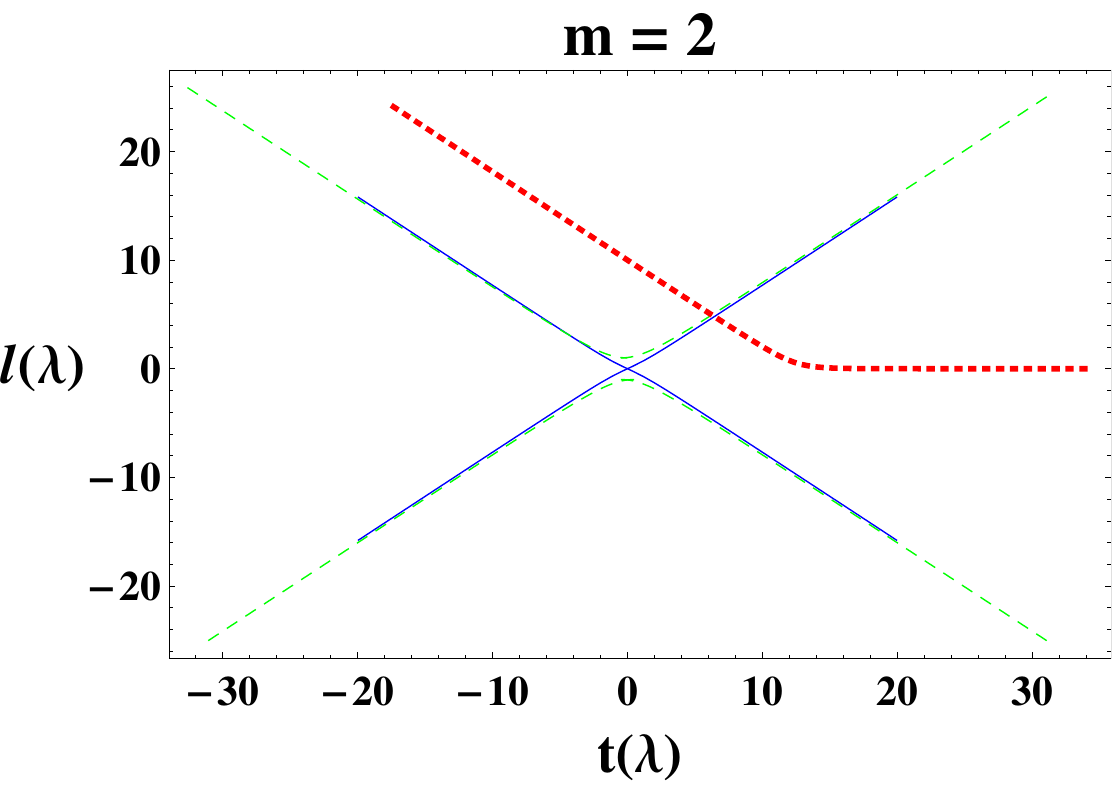}
\hspace{1cm}
\caption{Effective potential and geodesics for 4D-GEB wormhole ($m=2$): dotted-red, dashed-green, and continuous-blue curves representing -- Trapped, Returning and Crossing geodesics respectively.}
\label{fig:4d-trajectories-m2}
\end{figure} 

\begin{figure}[H]
\centering
\includegraphics[scale=.4]{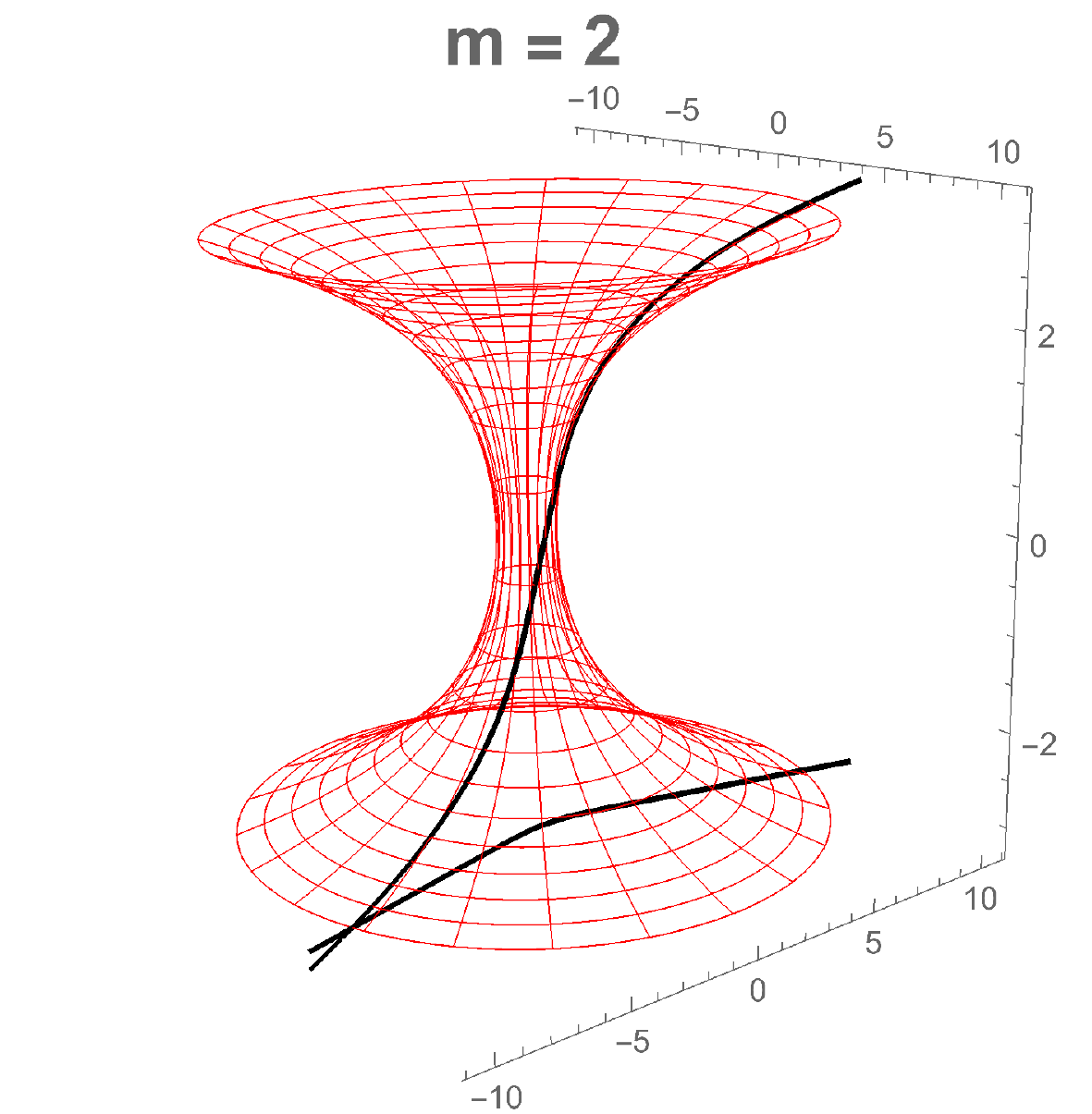}
\hspace{2cm}
\includegraphics[scale=.34]{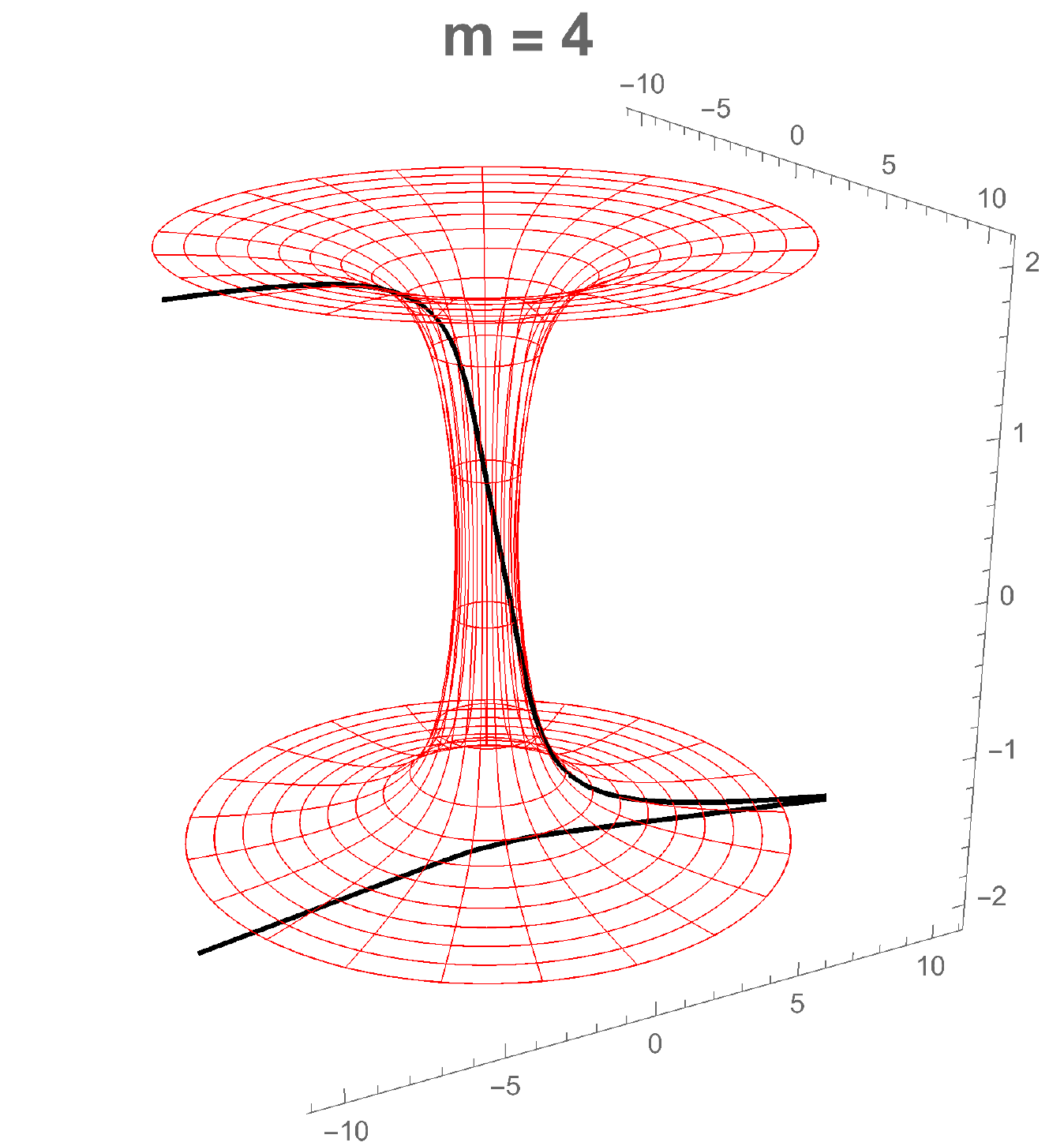}
\caption{Crossing geodesics in $m=2$ and $m=4$ 4D-GEB geometries}
\label{fig:Crossing-4D}
\end{figure}

In Fig. \ref{fig:Crossing-4D} we show the crossing geodesics for $m=2$ and $m=4$ in 4D-GEB geometries using the embedding diagrams discussed earlier. The curves at the bottom are the projections on $z=const.$ surfaces. In what follows, we shall not present such embedding diagrams for all the cases as we do not get any extra physical insight except aesthetically pleasant perspective.
 Below we present the solutions of the full geodesic equations for our 5D model.

\subsection{5D-WGEB model}

In general, for the 5D-WEB models, we get all three types of trajectories. The effective potentials, $V_{l}(\lambda)$ have similar profile for 4D and 5D models (with growing warp-factor) corresponding to all type of trajectories. However, as mentioned earlier, the existence of trapped trajectories is sensitive to the initial or boundary conditions in presence of the decaying warp-factor -- particularly the choice of $y(\l)$ at $\l=0$.
In 5D scenario, we also have the effective potential $V_{y}(\lambda)$ corresponding to the motion along the extra dimension as well. Note that here we only consider $m=2$ case as advantages of $m>2$ is provided by the warped extra dimension as demonstrated in \cite{Sharma:2021kqb}.

\begin{figure}[H]
\centering
\includegraphics[scale=.5]{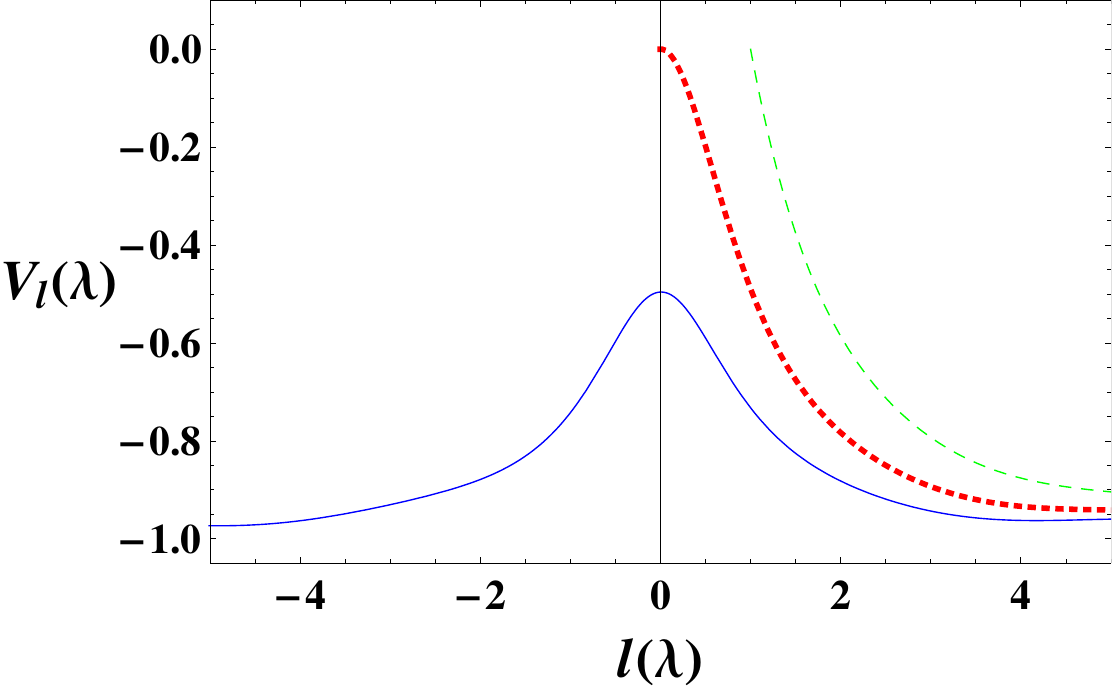}
\hspace{1cm}
\includegraphics[scale=.5]{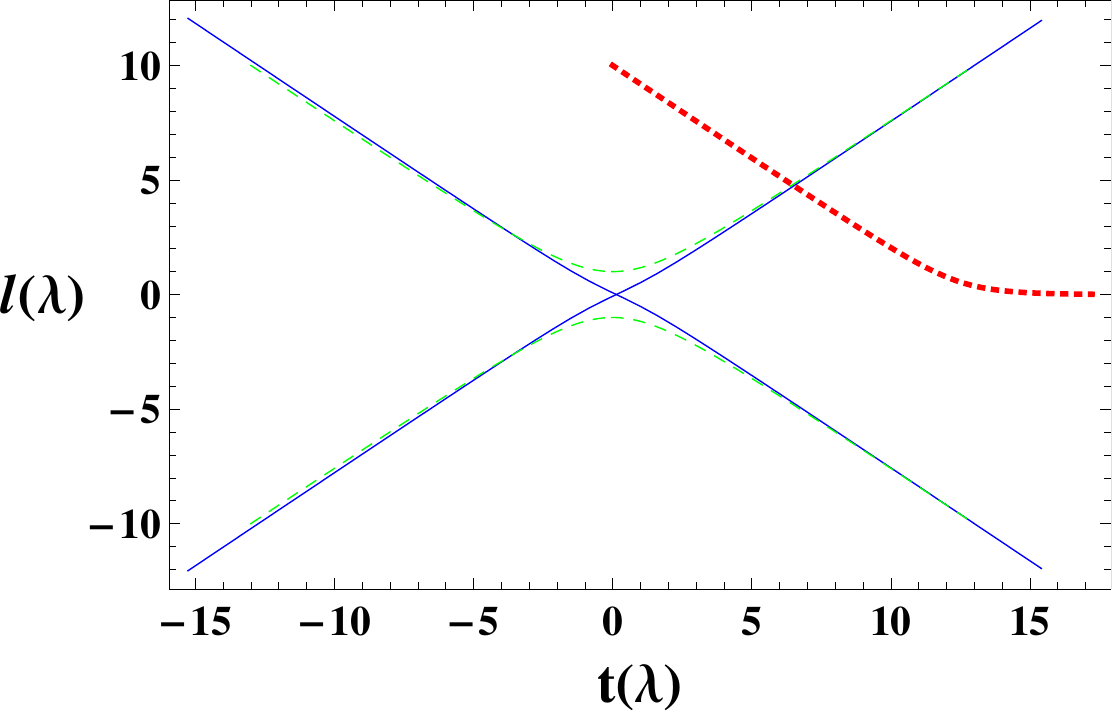}
\includegraphics[scale=.5]{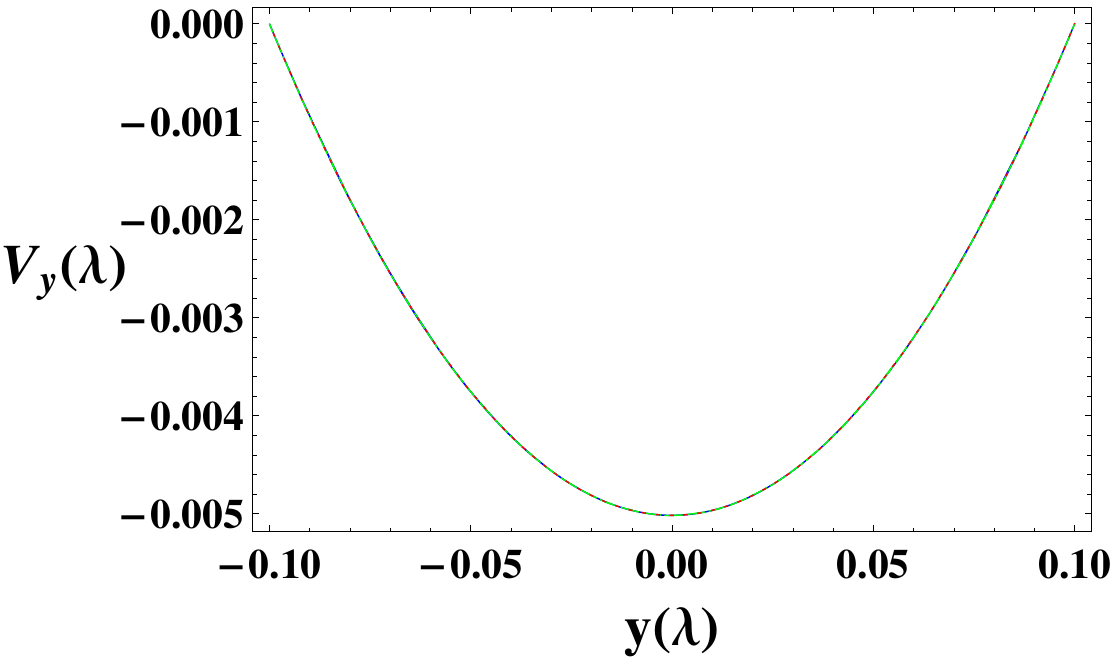}
\hspace{1cm}
\includegraphics[scale=.5]{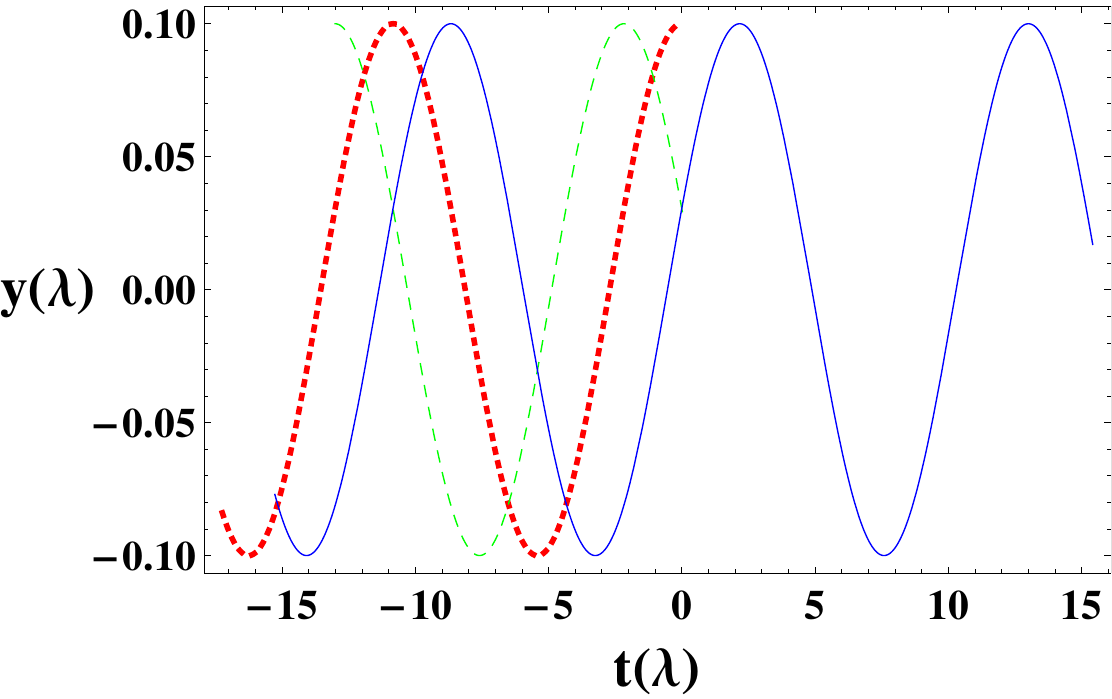}
\caption{Geodesics and effective potential for 5D-WEB model with growing warp factor : dotted-red, dashed-green, and continuous-blue curves representing -- Trapped, Returning and Crossing geodesics.}
\label{fig:5d-trajectories-growing}
\end{figure} 
Figure \ref{fig:5d-trajectories-growing} shows that in presence of the growing warp factor we have all three types of geodesics as per motion along $l$ is considered. Further, the trajectories are also confined near the location of the brane $y=0$ (as expected from the potential $V_{y}(\lambda)$ that looks like a potential for simple harmonic oscillator).
\begin{figure}[H]
\centering
\includegraphics[scale=.51]{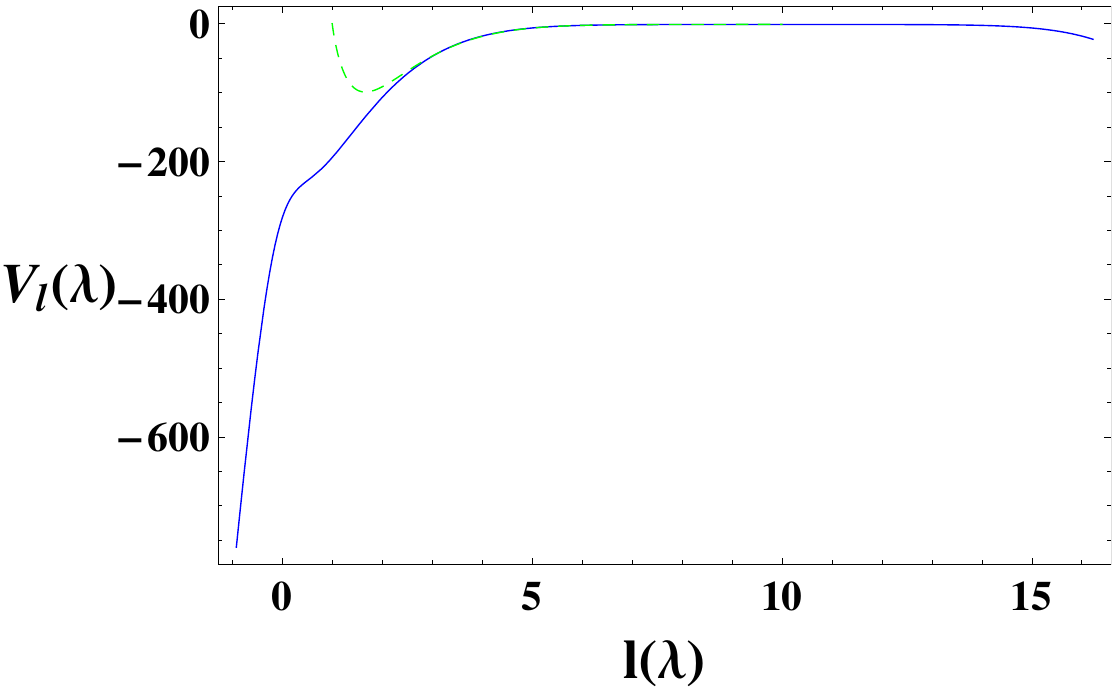}
\hspace{1cm}
\includegraphics[scale=.5]{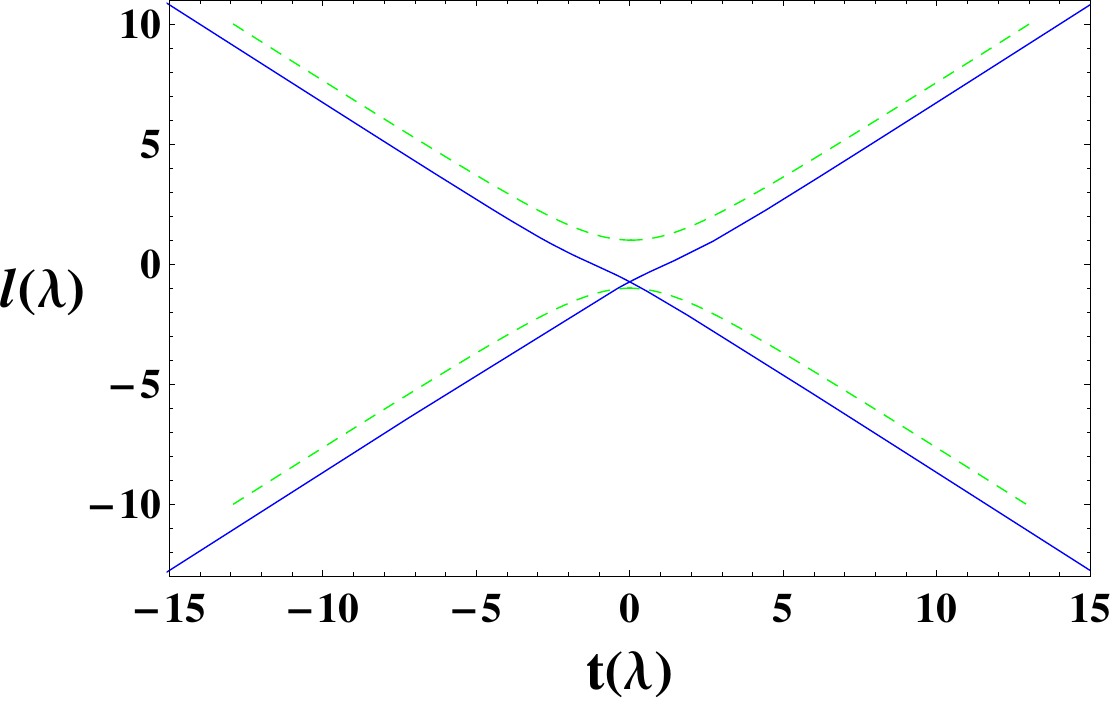}
\includegraphics[scale=.51]{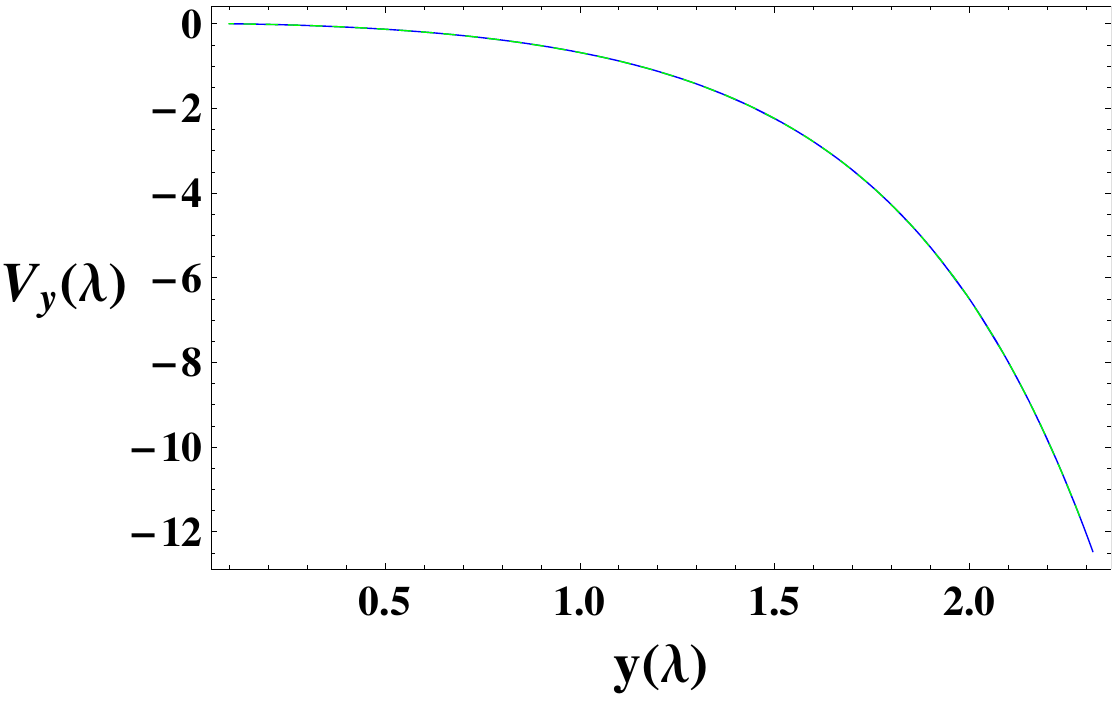}
\hspace{1cm}
\includegraphics[scale=.5]{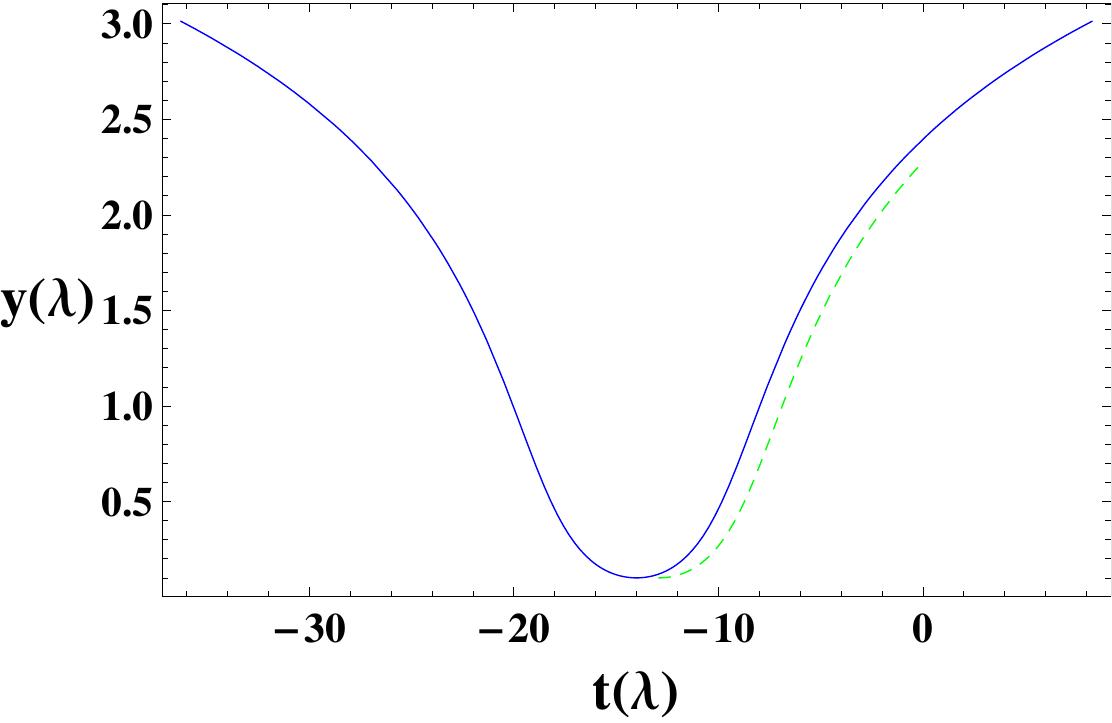}
\caption{Geodesics and effective potential for 5D-WEB model with decaying warp factor : dashed-green, and continuous-blue curves representing -- Returning and Crossing geodesics respectively.}
\label{fig:5d-trajectories-decaying}
\end{figure} 
  
On the other hand, we have runaway trajectories with decaying warp-factor as shown in fig. \ref{fig:5d-trajectories-decaying}). Here, given the boundary conditions we have used (see Appendix B), the trapped trajectories are absent. However, trapped trajectories exist even in the presence decaying warped factor very specific boundary condition.

\begin{figure}[H]
\centering
\includegraphics[scale=.5]{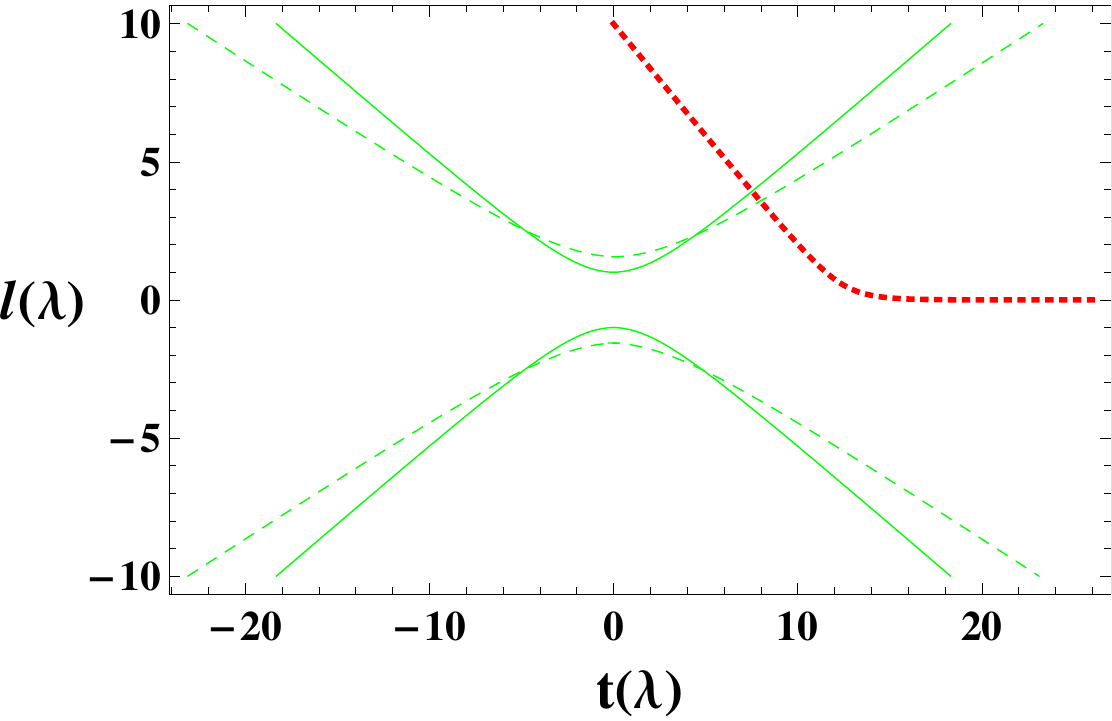}
\hspace{1cm}
\includegraphics[scale=.5]{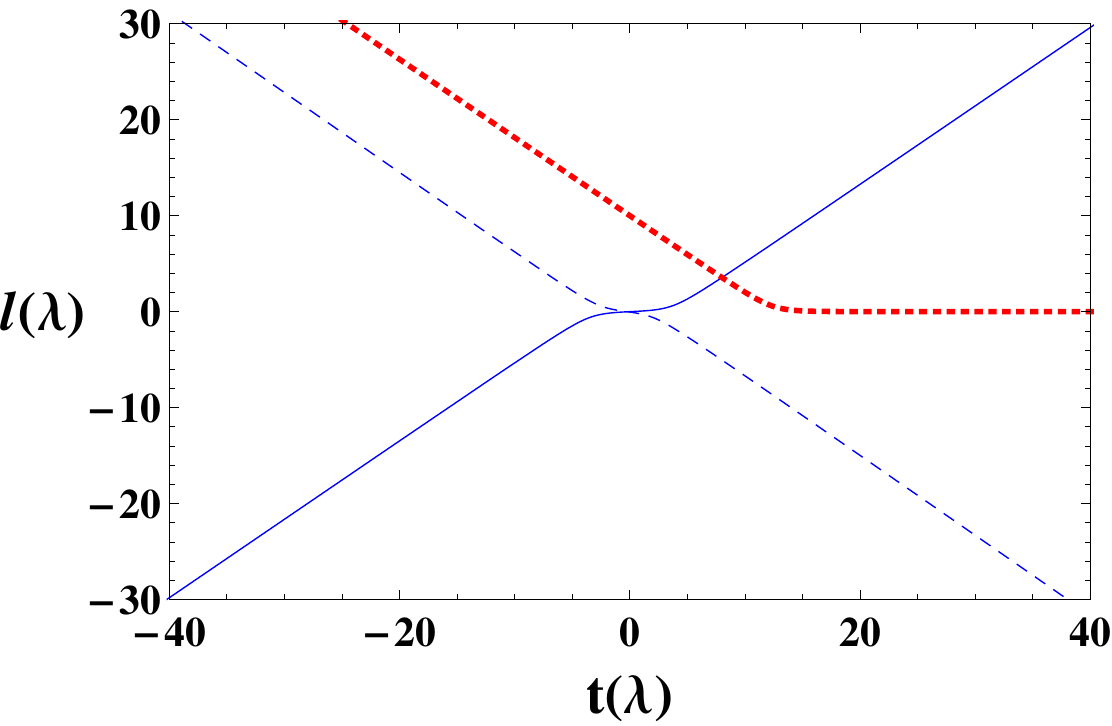}
\caption{Geodesics for 5D-WGEB model with growing (left) and decaying (right) warp factor. Here, $T = \sqrt{3}$, $H = \sqrt{2}$, $m = 2$, $b_{0} = 1$ and the dotted, continuous and dashed curves representing cases corresponding to three different initial values of $y(0)$ as mentioned in Appendix C. }
\label{fig:geodesics-effect-of-initial-y}
\end{figure}
Fig. \ref{fig:geodesics-effect-of-initial-y} shows that all three types of geodesics exist with suitably chosen boundary conditions. We set the value of $H = b_0\sqrt{T^{2} - 1}$ (which corresponds the condition for trapped trajectories in 4D or at $y=0$) and solve the geodesic equations for various values of $y(0)$. We find trapped trajectories corresponding to $y(0) = 0$ for both types of warp factors. Remarkably only returning and crossing trajectories exist corresponding to all other $y(0)$ values for growing (e.g. $y(0) = 0.881374, 1.01$) and decaying (e.g. $y(0) = 0.1, 0.3$) warp factors (see Appendix C for details of initial values used).             


\section{discussion} \label{sec:dis}

Wormholes necessitate the presence of exotic matter source (matter with a negative energy density), which is difficult to envisage on a macroscopic scale. To counter this problem, various models are proposed in the context of modified theories of gravity. Earlier we introduced a model where the (generalised) Ellis-Bronnikov wormhole is embedded in a 5D warped thick braneworld background and showed that corresponding energy conditions are satisfied.
In this work, we have explored the timelike particle trajectories in considerable detail, for both the 4D-GEB wormhole and 5D-WEB wormhole geometry. 
The results that reveal the effects of the wormhole parameter and the warping factor on the trajectories and are summarised below in a systematic manner.

\begin{itemize}

\item The embedding diagrams of the 4D-GEB geometry show that neck-length increases with increasing $m$. Notably, for 5D-WGEB model, in presence of the growing warp factor the neck length also increases with increasing $y$, whereas for decaying warp factor opposite effect is seen. The embedding diagrams also reveal the asymptotic flatness of both models.

\item The analytic approach reveals three possible types of particle trajectories for both 4D-GEB and 5D-WGEB wormhole models namely the trapped, returning and crossing trajectories given conditions (\ref{eq:4d-trajectory-condition}) and (\ref{eq:5d-trajectory-condition}) are satisfied. The general formula for the point of return $l_c$ is derived analytically. The amount of time the particle spends near the throat, depends on the value of $l_{c}$, which increases as $l_{c}$ decreases. A dynamical system analysis of the 4D geometry also provided confirmation for the existence of these trajectories.

\item In the case of trapped orbit-- a test particle begins free falling from infinity and spirals in asymptotically to the throat. These particles essentially orbit the throat for eternity. 
For returning trajectories-- freely falling test particles spirals in from infinity, never reach the throat $l = 0$ but return back to infinity from $l_{c}$.  
For crossing trajectories-- a test particle begins free falling from infinity on one side ($l \rightarrow \infty$), crosses the wormhole throat (may be after orbiting the throat multiple times depending on initial angular momentum), and then flies out to infinity on the other side ($l \rightarrow - \infty$). 

\item The geodetic potentials and all three types of particle trajectories for various boundary conditions are found through numerical evaluation and presented graphically. However, trapped geodesics are rare in presence of the decaying warp factor. The `neck-length' for the wormhole increases in presence of a growing warp factor whereas it decreases for a decaying warp factor as one moves further away from $y=0$.  

\item  In case of 5D-WGEB geometry, the effective potential $V_{y}(\lambda)$ is of oscillatory nature, for growing warp factor, implying that particles will be confined indefinitely around the location of the thick brane ($y = 0$). Whereas runaway trajectories (that disappears farther into the bulk) are observed in the case of decaying warp factor.

\end{itemize}

It will be interesting to investigate the congruence of timelike and null geodesics for these class of models and compare them to further understand the roles of the wormhole parameter and the warped extra dimension. The potential of WEB models as black hole mimickers would be an essential aspect to analyse. Stability of such wormhole geometry is an important issue to be addressed as well. One can further investigate the other astrophysical properties like deflection angle, lensing effect, photon sphere etc. We plan to report on these and more in future communications.




\section{Appendix}

\subsection{Appendix A}

\begin{center}
\begin{table}[!h]

\begin{adjustbox}{max width=\textwidth}
\begin{tabular}{|c|c|c|c|}
\hline
 \multicolumn{4}{|c|}{\textbf{Boundary Conditions used for Geodesics in 4D-GEB spacetime with $m = 2, 4, 6$}} \\ \hline
Variables & Trapped ($0 \leq \l \leq \l_{max}$) & Returning ($-\infty \leq \l \leq 0$) & Crossing ($-\infty \leq \l \leq \infty$) \\ \hline
$t(0)$& 0, 0, 0 & $\pm 31$, $\pm 12$, $\pm 12$ & 0, 0, 0 \\ 
$l(0)$ & 10, 10, 10 & $\pm 25$, $\pm 10$, $\pm 10$ & 0, 0, 0 \\ 
$\theta(0)$ & $\frac{\pi}{2}$, $\frac{\pi}{2}$, $\frac{\pi}{2}$ & $\frac{\pi}{2}$, $\frac{\pi}{2}$, $\frac{\pi}{2}$ & $\frac{\pi}{2}$, $\frac{\pi}{2}$, $\frac{\pi}{2}$ \\ 
$\phi(0)$ & 0, 0, 0 & 0, 0, 0 & 0, 0, 0 \\
$\dot{t}(0)$ & $\sqrt{3}$, $\sqrt{3}$, $\sqrt{3}$ & $\sqrt{3}$, $\sqrt{3}$, $\sqrt{3}$ & $\sqrt{3}$, $\sqrt{3}$, $\sqrt{3}$ \\
\hline 
\multicolumn{4}{|c|}{ $\dot{l}(0)$ is calculated from the geodesic constraint}    \\
 \hline
 $\dot{\theta}(0)$  & 0, 0, 0 & 0, 0, 0 & 0, 0, 0 \\
 $\dot{\phi}(0)$ & 0.0140021, 0.0141414, 0.0141421 & 0.00319489, 0.019999, 0.02 & 1, 1, 1 \\ \hline
\end{tabular}
\end{adjustbox}
\caption{\label{tab:table-4D} Boundary values used for Fig.(\ref{fig:4d-trajectories-m2}-\ref{fig:4d-trajectories-m6})}
\end{table}
\end{center}

\subsection{Appendix B}

\begin{center}
\begin{table}[!h]

\begin{adjustbox}{max width=\textwidth}

\begin{tabular}{ |c|c|c|c|  }
 \hline
 \multicolumn{4}{|c|}{\textbf{Boundary Conditions used for Geodesics in 5D-WEB (growing,decaying) spacetime}} \\
 \hline
 \textbf{Variables} & \textbf{Trapped} & \textbf{Returning} & \textbf{Crossing}\\
 \hline
 $t(0)$   & 0    &$\pm 13$, $\pm 3$  &  $ 13, \pm 14$      \\
 
 $l(0)$&   10  & $\pm 10$, $\pm 10$  & $\pm 10$, 10    \\
 
 $\theta(0)$ & $\frac{\pi}{2}$ &  $\frac{\pi}{2}$, $\frac{\pi}{2}$& $\frac{\pi}{2}$, $\frac{\pi}{2}$      \\
 
 $\phi(0)$   & 0 & 0, 0 &  0, 0     \\
 
 $y(0)$ &   0.1 & 0.1, 0.1 & 0.1, 0.1     \\
 
 $\dot{t}(0)$ & 1.71485  & $1.71485$, 1.74943   & $1.71485, 1.74943$ \\
\hline 
 
\multicolumn{4}{|c|}{ $\dot{l}(0)$ is calculated from the geodesic constraint}    \\
 
 \hline
 
 $\dot{\theta}(0)$ & 0 & 0, 0 & 0, 0      \\
 
 $\dot{\phi}(0)$ & $0.0138282$ & $0.019556$, 0.0200503 & 0.00977802, 0.0100251\\
 
 $\dot{y}(0)$ & 0 & 0, 0 & 0, 0\\
 
 \hline

\end{tabular}
\end{adjustbox}

\caption{\label{tab:table-5D} Boundary values used for Fig.\ref{fig:5d-trajectories-growing} and \ref{fig:5d-trajectories-decaying}}
\end{table}

\end{center} 


\subsection{Appendix C}

\begin{center}

\begin{table}[!h]

\begin{adjustbox}{max width=\textwidth}

\begin{tabular}{ |c|c|c|c|  }
 \hline
 \multicolumn{4}{|c|}{\textbf{Boundary Conditions for Geodesics with various values of $y(0)$ (growing,decaying)}} \\
 \hline
 \textbf{Variables} & \textbf{(dotted-curves)} & \textbf{(continuous-curves)} & \textbf{(dashed-curves)}\\
 \hline
 $t(0)$   & 0, 0    &$\pm 18.3$, 16 &   $\pm 23.3$, 14      \\
 
 $l(0)$&   10, 10  & $\pm 10$, 10  & $\pm 10$, $- 10$    \\
 
 $\theta(0)$ & $\frac{\pi}{2}$, $\frac{\pi}{2}$ &  $\frac{\pi}{2}$, $\frac{\pi}{2}$& $\frac{\pi}{2}$, $\frac{\pi}{2}$      \\
 
 $\phi(0)$   & 0, 0 & 0, 0 &  0, 0     \\
 
 $y(0)$&   0, 0 & 0.881374, 0.1 & 1.01, 0.3     \\
 
 $\dot{t}(0)$ & $\sqrt{3}$, $\sqrt{3}$  & 0.866025, 1.74943   & 0.716391, 1.89267 \\

\hline 
 
\multicolumn{4}{|c|}{ $\dot{l}(0)$ is calculated from the geodesic constraint}    \\
 
 \hline
 
 $\dot{\theta}(0)$ & 0, 0 & 0, 0 & 0, 0      \\
 
 $\dot{\phi}(0)$ & 0.0140021, 0.0140021 & 0.00700105, 0.0141426 & 0.00579139, 0.0153006\\
 
 $\dot{y}(0)$ & 0, 0 & 0, 0 & 0, 0\\
 
 \hline

\end{tabular}
\end{adjustbox}

\caption{\label{tab:table-y}Boundary values used for Fig.(\ref{fig:geodesics-effect-of-initial-y})}
\end{table}

\end{center} 







\end{document}